\documentclass[10pt,sigconf,letterpaper,nonacm]{acmart}

\usepackage{amsmath, amsfonts}
\usepackage{siunitx}

\usepackage[english]{babel}
\usepackage{indentfirst}
\usepackage{paralist}   % compactitem 等
\usepackage{enumitem}   % 控制 itemize 间距
\usepackage{multicol}
\usepackage{comment}
\usepackage{subfiles}
\usepackage{blindtext}
\usepackage{textcomp}
\usepackage{mdframed}
\usepackage{fancybox}
\usepackage{framed}

\usepackage{xcolor}

\newcommand{\bd}[1]{\textit{#1}}

\usepackage{graphicx}
\usepackage{float}
\usepackage{subfigure}
\usepackage{tikz}
\usetikzlibrary{arrows.meta, automata, positioning, shapes.callouts}
\usepackage{pgfplots}
\pgfplotsset{compat=1.18}
\usepackage{pgf-pie}
\usetikzlibrary{patterns}

\usepackage{booktabs}
\usepackage{multirow}
\usepackage{threeparttable}
\usepackage{tabularx}
\usepackage{makecell}
\usepackage{colortbl}
\usepackage[table]{xcolor}
\usepackage{arydshln}   % \hdashline
\newcolumntype{L}{>{\arraybackslash}X}

\usepackage{pifont}
\usepackage{bbding}

\usepackage{algorithm}
\usepackage[noend]{algpseudocode}
\algrenewcommand\algorithmicrequire{\textbf{Input:}}
\algrenewcommand\algorithmicensure{\textbf{Output:}}

\usepackage[font=small,labelfont=bf,tableposition=top]{caption}
\usepackage{listings}
\usepackage{xcolor}

\lstdefinestyle{trace}{
    basicstyle=\ttfamily\small,
    breaklines=true,
    breakatwhitespace=true,
    columns=fullflexible,
    frame=single,
    framerule=0.3pt,
    xleftmargin=1em,
    xrightmargin=1em,
    backgroundcolor=\color{gray!3},
}

\theoremstyle{plain}

\theoremstyle{definition}

\usepackage{todonotes}
\usepackage{trace}
\usepackage{filecontents} % 如果真要 inline 建 bib；否则可以删
\def\BibTeX{{\rm B\kern-.05em{\sc i\kern-.025em b}\kern-.08em
    T\kern-.1667em\lower.7ex\hbox{E}\kern-.125emX}}

\usepackage{url}

\makeatletter
\def\Url@Underline{}
\makeatother

\newcommand{\addr}[2]{\href{#2}{\texttt{#1}}}
\newcommand{\hlhref}[2]{\href{#1}{\textcolor{teal}{#2}}}

\AddToHook{begindocument/end}{%
  \hypersetup{
    colorlinks=true,
    linkcolor=blue,
    urlcolor=magenta,
    citecolor=magenta,
  }%
}

\newenvironment{packeditemize}{
	\begin{list}{$\bullet$}{
			\setlength{\labelwidth}{4pt}
			\setlength{\itemsep}{0pt}
			\setlength{\leftmargin}{\labelwidth}
			\addtolength{\leftmargin}{\labelsep}
			\setlength{\parindent}{0pt}
			\setlength{\listparindent}{\parindent}
			\setlength{\parsep}{0pt}
			\setlength{\topsep}{1pt}}}{\end{list}}

\newcommand{\imcpapernumber}{131}

\acmYear{2026}
\copyrightyear{2026}
\setcopyright{none}
\acmSubmissionID{Paper \#\imcpapernumber\,, \pageref{TotPages}\ pages}
\acmConference[IMC '26]{ACM Internet Measurement Conference}{October 12--16, 2026}{Karlsruhe, Germany}

\begin{document}
\title{MEV in Binance Builder}

%=================================================
%author
%=================================================

\author{$^{1,3}$Qin Wang, $^2$Ruiqiang Li, $^3$Guangsheng Yu, $^4$Vincent Gramoli, $^1$Shiping Chen}
\affiliation{%
\vspace{5pt}
  \institution{\textit{$^1$CSIRO Data61} $|$ \textit{$^2$University of Wollongong} \\ \textit{$^3$University of Technology Sydney}  $|$ \textit{$^4$University of Sydney}}
  \country{Sydney, Australia}
}

%=================================================
%abstract
%=================================================

\begin{abstract}
We study builder-driven MEV arbitrage on BNB Smart Chain (BSC). BSC’s Proposer-Builder Separation (PBS) adopts a leaner design: only whitelisted builders can participate, blocks are produced at shorter intervals, and private order flow bypasses the public mempool. These features have long raised community concerns over centralization, which we empirically confirm by tracing the arbitrage activities of the two dominant builders from Apr.~1,~2025 to Feb.~28,~2026 (full observable activity cycle). Within months, the two leading builders, \bd{48Club} and \bd{Blockrazor}, produced over 87\% of blocks and captured about 90\%+ of MEV profits. 

We find that profits concentrate in short, low-hop arbitrage routes over wrapped tokens and stablecoins, and that block construction rapidly converges toward monopoly. Beyond concentration alone, our analysis reveals a structural source of inequality: BSC’s short block interval and whitelisted PBS collapse the contestable window for MEV competition, amplifying latency advantages and excluding slower builders and searchers. MEV extraction on BSC is not only more centralized than on Ethereum, but also structurally more vulnerable to censorship and fairness erosion.

\end{abstract}


%=================================================
\maketitle
%=================================================   

%=================================================
\section{Introduction}
\label{sec-intro}
%=================================================   

Maximal Extractable Value (MEV)~\cite{daian2020flash,qin2022quantifying} refers to the profit that can be extracted by reordering, inserting, or censoring transactions within a block~\cite{zhou2023sok,heimbach2022sok}. To manage MEV extraction more fairly, Ethereum introduced the Proposer-Builder Separation (PBS) model~\cite{heimbach2023ethereum} to decouple block proposal from block construction. In PBS, specialized entities called \textit{builders} are responsible for assembling profitable blocks, while \textit{proposers} (or validators) select the highest-bidding block to propose. PBS aims to reduce validator centralization~\cite{grandjean2024ethereum,yang2025decentralization} and democratize MEV opportunities by introducing a competitive builder marketplace~\cite{wu2024compete,oz2024wins,wu2024strategic}.

\begin{figure}[t]
    \centering
    \subfigure[MEV block counts by builder over time. 48Club and Blockrazor dominate throughout the observation window.]{
        \includegraphics[width=\linewidth]{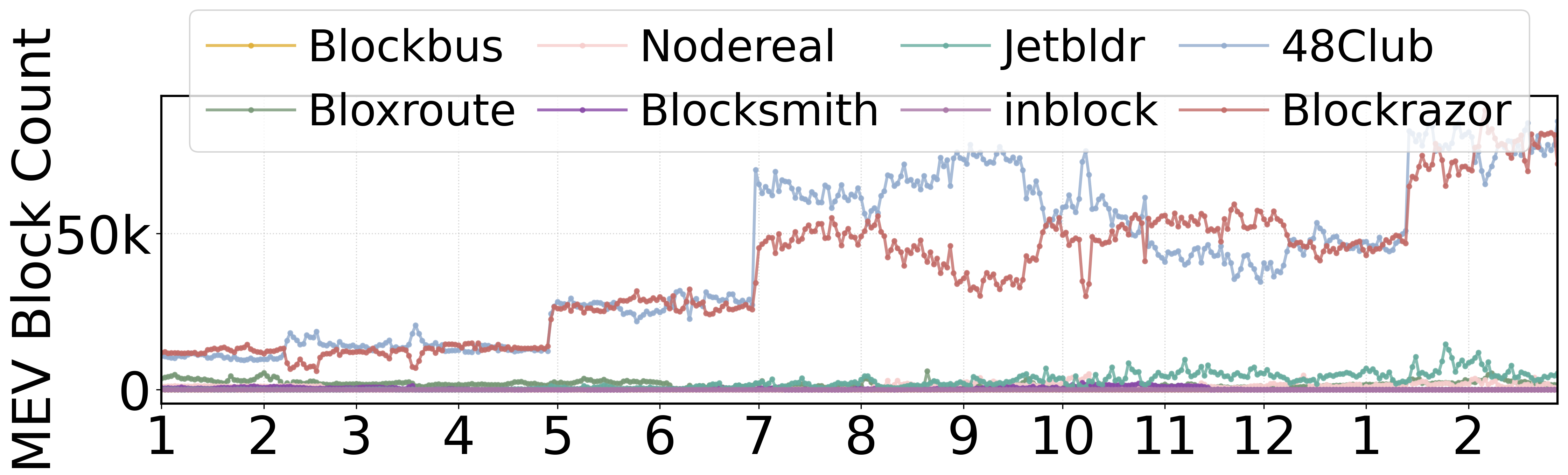}
        \label{fig:bsc_block}
    }
    \hfill
    \subfigure[Builder market share over time. The top two builders consistently account for more than 87\% of blocks.]{
        \includegraphics[width=\linewidth]{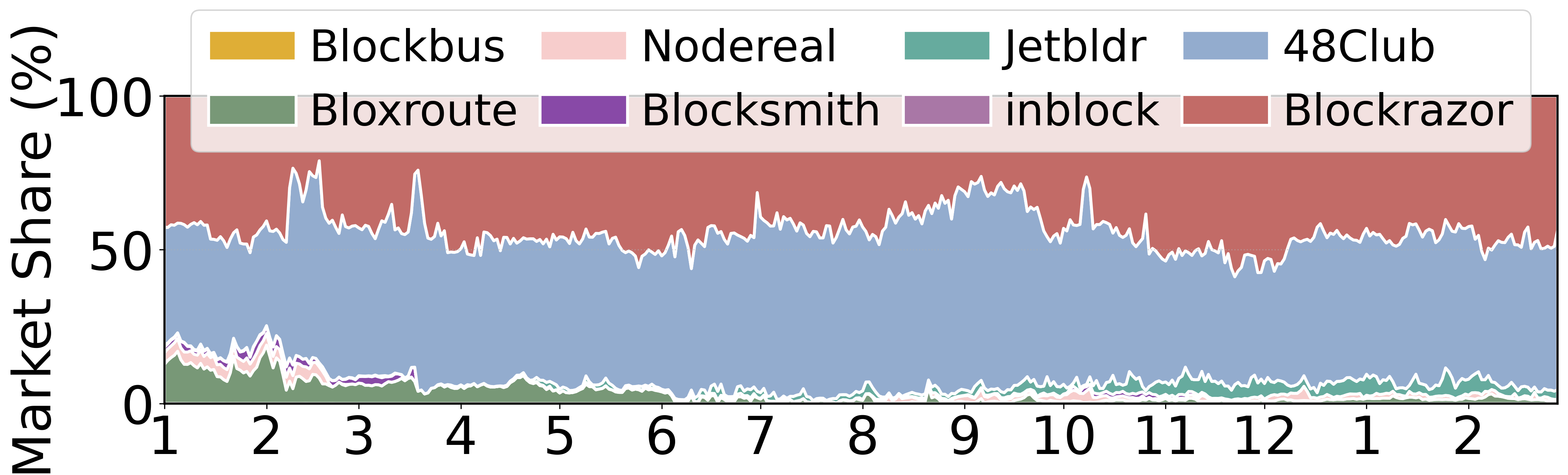}
        \label{fig:bsc_share}
    }
    \caption{Builder activity and market concentration on BSC.}
    \label{fig:builderMarket}
\end{figure}

\begin{table*}[t]
\centering
\footnotesize
\caption{Comparison of PBS adoption across chains.}
\label{tab:pbs-comparison}
\resizebox{0.9\linewidth}{!}{
\begin{tabular}{c |c |c c |c| c }
\toprule
\textbf{Chain} & \textbf{PBS Status} & \textbf{Builders} & \textbf{Top-2 Share} & \textbf{Relay}   & \textbf{Block Time}  \\
\midrule
\textbf{Ethereum} & Live (MEV-Boost) 
  & 20--30 active & 85--90\% 
  & mandatory relays & 12s \\

\rowcolor{pink!50} \textbf{BSC} & Live (BEP-322) 
  & 9 brands (35 inst.) & $>$87\% 
  & validator-direct &  3s  \\

\textbf{Polygon PoS} & Pilot (Flashbots relay) 
  & (a few) trial builders & n/a & external relays & $\sim$2s~\cite{polygon2025block} \\

\textbf{Avalanche C-Chain} & Early Tooling \cite{avalanche2025hub} & n/a & n/a & no formal PBS & $\sim$1.8s \cite{avalanche2025block} \\

\textbf{Solana} & no PBS, 
  built-in priority fees & n/a & n/a & no relays & $\sim$400ms \cite{solana2025block}  \\
\bottomrule
\end{tabular}
}
\end{table*}

BNB Smart Chain (BSC) adopted the PBS model~\cite{bnbpbs1,bnbpbsNews} in early 2024 via BEP-322~\cite{bnbpbs}. Compared to Ethereum, BSC's implementation introduces differences (more in \S\ref{subsec-compare}):

\begin{packeditemize}
    \item only a small whitelisted set of builders, admitted through the Builder API, are eligible to construct blocks~\cite{bnbpbsProxy};
    \item faster block times (3s under BSC’s Proof of Staked Authority fast-finality design~\cite{bep126,everything,li2025finality});
    \item builders maintain direct relationships with validators in a validator set of only about 40 entities, operating without relays or public block auctions, and support zero-gas-price transactions (0 Gwei) that create private order flow through MEV-protected and builder-operated RPC endpoints instead of gas-fee competition~\cite{yang2025decentralization,wu2024compete}.
\end{packeditemize}

This architecture provides builders with advantages: (i) BSC allows private transaction submission via MEV-protected RPC endpoints, forming a private order flow channel between users (including searchers) and builders \cite{wang2025private,lyu2025demystifying}. These transactions are invisible to the public mempool \cite{weintraub2022flash}, giving builders (and any exclusive partners) early access to capture MEV opportunities~\cite{duneData}. (ii) BSC's short block time limits the reaction window for external searchers, who must deliver bundles within tight latency constraints (often less than 200ms) to be considered for inclusion. Builders, who control the block, can directly attach last-moment arbitrage transactions without competition. (iii) The 0 Gwei fee policy incentivizes the use of copy-trading and sandwich attack strategies, as transactions can be duplicated around a target with negligible cost.

As a result, the builder market on BSC has become highly concentrated. Our investigation (cf. raw block counts and their relative block-construction share in Fig.\ref{fig:builderMarket} and Table~\ref{tab:builder}) shows that the two leading builders (i.e., \bd{48Club}, \bd{Blockrazor}) produce a little over 87\% of all blocks, while the remaining seven builders account for only about 12\%. This creates an asymmetry: MEV searchers and ordinary users lack direct access to block construction and must instead submit bundles to a handful of dominant builders, effectively centralizing transaction visibility and ordering power.

Community discussions echo these concerns. For example, user “cracksparrow00” \cite{bnbdiscuss1,bnbdiscuss2} described profitable arbitrage through private order flow and argued that builders now capture 90--95\% of MEV. While BscScan alone does not reveal execution identity, the recurring addresses and scale of observed profits strongly suggest builder-linked self-arbitrage rather than ordinary searcher activity. Unless otherwise stated, all profit figures refer to net profit after subtracting redistributed share payouts.

In this paper, we present the first systematic, empirical on-chain study of builder-driven MEV on BSC. In particular,

\vspace{3pt}
\begin{packeditemize}
    \item we trace and analyze arbitrage contracts deployed by the two dominant builders, uncovering consistent use of short-hop (e.g., two--three swap) stablecoin routes and profit margins unattainable by ordinary searchers;
    \vspace{3pt}
    \item we quantify builder concentration at scale, showing that \bd{48Club} and \bd{Blockrazor} produce over 87\% of blocks, while \bd{48Club} alone retains about 75\% of the net profit observed in our two-builder dataset;
    \vspace{3pt}
   \item we provide a curated dataset covering 9.63M builder-executed arbitrage transactions across 250+ factories and 23{,}000+ pools, with detailed routing and profit-attribution information.
   \vspace{3pt}
    \item we identify a structural source of MEV inequality in BSC’s PBS design, showing how short block intervals and whitelisted builders collapse the contestable window for MEV competition.
\end{packeditemize}

\vspace{3pt}
We do not claim novelty in MEV optimizations or mitigations. We send our salute to the line of prior empirical studies on MEV. An incomplete list includes \cite{yang2025decentralization,wang2025private,gerzon2025quantifying,wu2025measuring,ferreira2024rolling,qin2022quantifying}.

%=================================================
\section{Binance's PBS}
\label{sec-bck}
%=================================================

We present BSC (\S\ref{subsec-bscdesign}) and PBS design (\S\ref{subsec-bscpbs}, \S\ref{subsec-arbitrage}) and contrast it with Ethereum PBS (\S\ref{subsec-compare}). We further summarize key PBS designs across major chains (Table~\ref{tab:pbs-comparison}).

\subsection{BSC Design}
\label{subsec-bscdesign}

BSC is an EVM-compatible blockchain designed to provide low-latency finality and high throughput while retaining close compatibility with the Ethereum execution model. 
Its consensus protocol, Proof of Staked Authority (PoSA)~\cite{bep126,li2025finality}, combines delegated staking with a rotating validator set. 
A fixed number of validators (typically 40–45) are elected for each epoch and take turns proposing blocks in a round-robin fashion every 3 seconds. 
Validators must self-bond substantial amounts of BNB and can be slashed or replaced if they miss blocks or misbehave, creating strong economic and reputational incentives for timely block production.

\subsection{BSC PBS}
\label{subsec-bscpbs}

\noindent\textbf{Architecture and roles.}
BSC’s PBS framework separates block economics from consensus. Under PoSA, a small set of stake-bonded validators rotate every 3~seconds as proposers, while external builders specialize in the block’s \emph{economic content}: transaction selection, ordering, arbitrage routes, and MEV bundle composition. Proposers mainly check consensus validity, including gas usage, nonces, state-root correctness, and block execution, before signing the block.

This setting has two structural implications. First, the 3-second slot leaves little room for multi-round commit--reveal, resulting in a single-round builder--proposer interaction. Second, because BSC builders do not independently compute state transitions, proposers must execute the full transaction list and derive the canonical block header. Together, these constraints create a \emph{direct builder--proposer pipeline} without relays. Through BEP-322~\cite{bnbpbs}, builders communicate with the scheduled proposer by submitting bids that include, or later deliver, full transaction lists with optional MEV bundles.

We next show the workflow induced by these constraints.

\begin{figure}[t]
    \centering
    \subfigure[BSC PBS: whitelisted builders submit bids directly to proposers, who then execute the full transaction list and finalize the block in a single-round bid-and-block handoff.]{
        \includegraphics[width=0.99\linewidth]{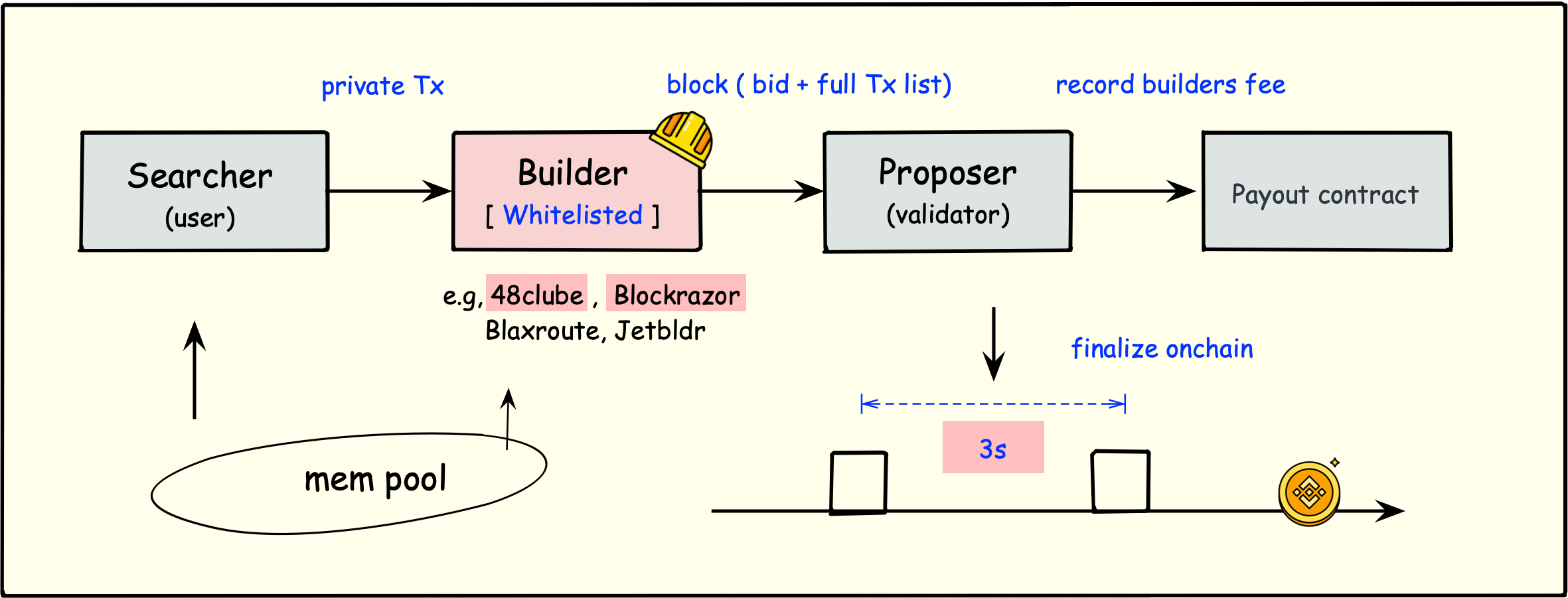}
        \label{fig:bsc_pbs}
    }
    \hfill
    \subfigure[Ethereum PBS: permissionless builders bid through relays, proposers sign only the block header first, and the full block is revealed afterward through a relay-mediated commit--reveal workflow.]{
        \includegraphics[width=0.99\linewidth]{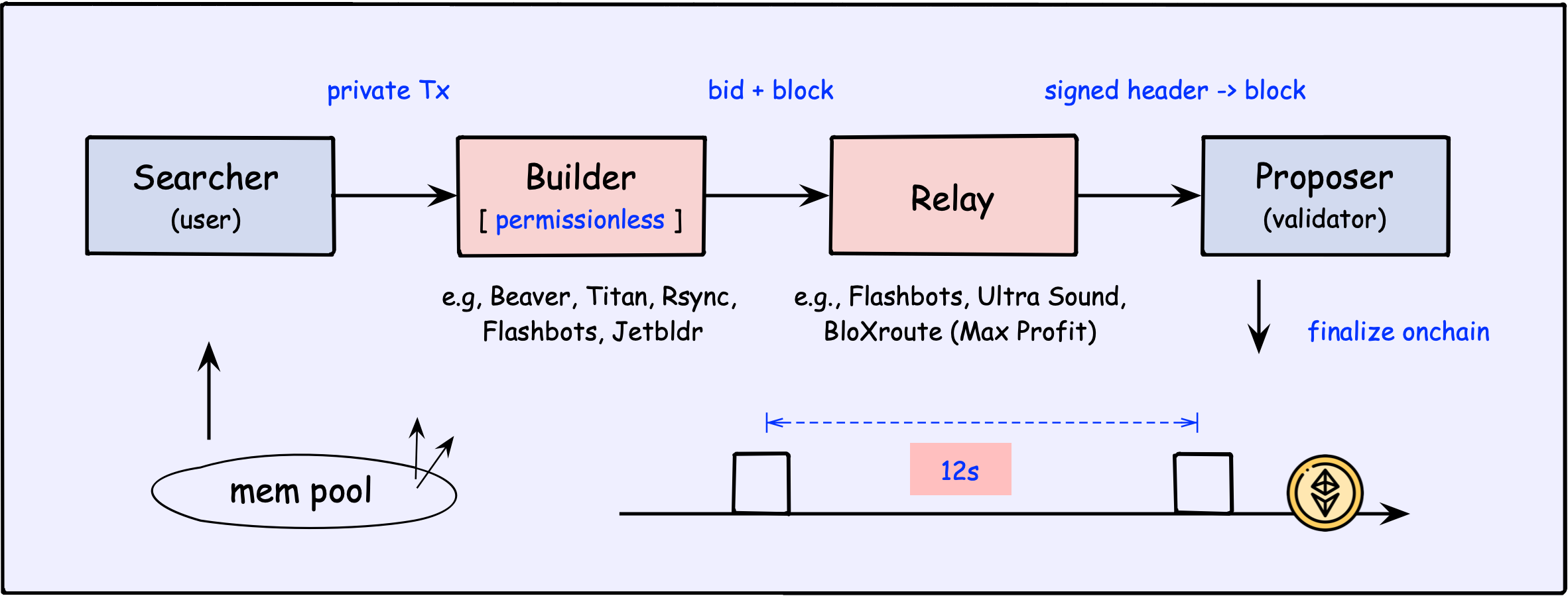}
        \label{fig:eth_pbs}
    }
    \caption{Comparison of two PBS workflows.}
    \label{fig:pbs_comparison}
\end{figure}

\vspace{3pt}
\noindent\textbf{How to produce a block.}
Each block interval proceeds through the following steps
(Fig.~\ref{fig:bsc_pbs}):

\begin{packeditemize}
    \item \textcolor{teal}{\underline{i)} Bid submission:}
        Builders submit compact bids containing the height, timestamp, Merkle
        root of transactions, expected gas fees, and offered payment.
    \item \textcolor{teal}{\underline{ii)} Bid selection:}
        Proposer selects the highest-paying bid.
    \item \textcolor{teal}{\underline{iii)} Receipt:}
        The proposer issues a signed receipt and requests the full
        transaction list.
    \item \textcolor{teal}{\underline{iv)} Block delivery:}
        The builder provides the full block body (or embeds it within the bid
        in one-step mode).
    \item \textcolor{teal}{\underline{v)} Validation:}
        The proposer locally executes the block, verifies correctness, and
        signs it.
\end{packeditemize}

If the builder fails to respond or provides invalid contents, the proposer rejects the bid, locally blacklists the offending builder, and moves to the next-best bid. In the absence of a valid bid, the proposer falls back to building a block
from its local mempool. Such fallbacks were extremely rare in our dataset (<0.02\%), indicating stable builder infrastructure.

\vspace{3pt}
\noindent\textbf{MEV extraction and distribution.}
Builders consolidate public and private order flow, reorder transactions, and bid for block inclusion. To preserve BSC’s gas-fee sharing model (fees distributed to all validators), BEP-322 requires proposers to insert a zero-gas accounting
transaction recording the builder’s payment in a payout contract; builders later claim these rewards on-chain. Misbehavior (e.g., non-delivery after winning a bid) is discouraged via signed receipts, reputational consequences, and local blacklisting.

% --------------------------------------------------------------------
\subsection{How to Arbitrage?}
\label{subsec-arbitrage}

Builders derive most of their bidding advantage from \emph{internal arbitrage}.
Let $\mathcal{S}=\{s_1,\dots,s_k\}$ denote a proprietary strategy set, where
each $s_i$ encodes routing heuristics, latency constraints, and bundle rules.
Given a transaction stream $\mathcal{T}$, a strategy $s$ extracts candidate
opportunities $O=f_s(\mathcal{T})$. An optional evasion operator
$\mathcal{E}$ may transform $O$ into $O'$ to mitigate frontrunning:
\[
\mathcal{T}\xrightarrow{\,f_s\,}O\xrightarrow{[\mathcal{E}]}O'
\xrightarrow{\textsc{ArbitrageRun}}\Delta.
\]

Consider a typical three-hop path executed by builders (aligned with the example in Appendix~A):
\[
\text{USDT}\rightarrow\text{WBNB}\rightarrow\text{USD1}\rightarrow\text{USDT}.
\]
From the AMM trace, the builder reconstructs:
\begin{packeditemize}
    \item $T$: the sequence of on-chain events (swaps, syncs, transfers),
    \item $P$: the ordered multi-hop path,
    \item $\Phi$: the realized profit flow (input, output, redistribution),
    \item $\Lambda$: deductions such as validator share-profit or gas.
\end{packeditemize}

Algorithm~\ref{algo-extractArbi} shows how $P$ is extracted from $T$. 
Algorithm~\ref{algo-attributeprofit} computes $(\Phi,\Lambda)$ to obtain net profit. We keep the
main-text description informal and defer pseudocode to Appendix~\ref{apd-example}.

\vspace{3pt}
\noindent\textbf{Formal path representation.}
An arbitrage opportunity $O'$ is represented as:
\[
\big(\langle T_0,\ldots,T_n\rangle,\;
  \langle P_1,\ldots,P_n\rangle,\;
  \langle F_1,\ldots,F_n\rangle,\;
  \langle D_1,\ldots,D_n\rangle \big),
\]
where $T_i$ denotes tokens, $P_i$ pools, $F_i\!\in\!\{0,1\}$ pool-type flags (V2/V3), and $D_i\!\in\!\{0,1\}$ swap-direction flags. This tuple provides the \emph{execution blueprint} for a multi-hop arbitrage: it specifies the assets, pools, and directions required to realize the opportunity and serves as the internal object that builders evaluate before
embedding a path into a block. An implementation template is provided in Appendix~\ref{apd-imple}.

In the remainder of the paper, we use this abstraction implicitly: each observed arbitrage transaction corresponds to an instantiated path descriptor, whose realized profit and redistribution are measured empirically.

\vspace{3pt}
\noindent\textbf{Builder-side execution pipeline.}
At block time, builders use this path descriptor as the basis for execution:
they evaluate candidate opportunities derived from public and private
order flow, simulate multi-hop routes consistent with the tuple above, and embed
the chosen opportunities into their block payloads. Only the resulting swaps
and redistribution transfers appear on-chain, which is why our reconstruction
does not include builder. AMM events alone encode sufficient
structure to recover the executed cycle.

\subsection{Compared with Existing PBS (Fig.\ref{fig:pbs_comparison})}
\label{subsec-compare}

We mainly compare BSC’s direct builder--proposer scheme with Ethereum’s relay-mediated PBS.

\begin{packeditemize}
\item\textit{Removal of trusted relay.}  
Ethereum’s PBS (via MEV-Boost) introduces relays because proposers are numerous and mutually untrusted. Builders submit sealed block headers to a relay, which forwards only the header and bid to the proposer; the proposer signs the header without seeing the block body, and the relay reveals the full block afterward. This prevents proposers from stealing order flow but adds complexity and centralizes trust in relays. BSC uses no relay. Builders send bids directly to the proposer, and reveal the full transaction list only after receiving a signed receipt. With a small, stake-bonded validator set ($\sim$40), the system relies on proposer reputation and the receipt mechanism rather than commit–reveal.
    
\item\textit{Block proposal.}  
Ethereum builders assemble full blocks and expose only the header to proposers, who finalize it blindly.  
In BSC, builders cannot compute the canonical state root; after a bid is accepted, they submit the full transaction list, and the proposer executes all transactions, derives the final header, and signs the block. Block contents are therefore visible before finalization.

\item\textit{Block intervals.}  
BSC’s 3-second block interval leaves little room for multi-round protocols, favoring a single-round bid and block handoff.  
Ethereum’s 12-second slot allows for the additional relay round trip and stronger secrecy assumptions.  
BSC thus opts for a more direct workflow, whereas Ethereum prioritizes trust minimization even at the cost of latency.

\item\textit{Rewards.}  
Ethereum builders set as coinbase and include a payment to the proposer, splitting MEV within the block.  
BSC cannot reassign coinbase rewards due to delegated staking. Instead, builders are compensated through an explicit on-chain payment contract, while proposers retain standard block rewards for distribution to delegators.

\end{packeditemize}

\vspace{3pt}
We briefly discuss other PBS deployments (Table~\ref{tab:pbs-comparison}). Polygon PoS experiments with Flashbots-style relay builders, Avalanche provides early tooling without a sealed-header protocol, and Solana relies on priority fees instead of PBS.

%=================================================
\section{Our Measurement}
\label{sec-method}
%=================================================

%\subsection{Methodology}
%\label{subsec:method}

Our empirical analysis directly instantiates the arbitrage abstraction introduced in \S\ref{subsec-arbitrage}. Specifically, each executed arbitrage cycle is reconstructed as a concrete realization of the path tuple $\langle T, P, F, D \rangle$, while the reported profits correspond to the realized $\Delta$ after accounting for deductions $\Lambda$.

Following this abstraction, our measurement proceeds in two steps. First, we survey all active builders on BSC and quantify their block production shares (\S\ref{subsec-builder}). This reveals a highly skewed distribution, with only a small subset of builders producing the vast majority of blocks. We therefore focus our analysis on this top group, which accounts for 87\% of builder-produced blocks. Second, we examine the on-chain data associated with these builders (\S\ref{subsec-dataset}), concentrating on arbitrage-related transactions. We analyze the scale of their activity, token usage, path structures, and profit attribution, allowing us to assess both the concentration of power and the mechanisms through which MEV is extracted. 

Appendix~\ref{apd-data} further details these procedures with concrete examples and pseudocode.

\subsection{Data Collection}
\label{subsec:data-collection}

We collect builder-driven MEV activities by full-block traces, transaction receipts, and contract-level metadata. 

Our collection pipeline continuously fetches: (i) block headers and bodies from BSC RPC nodes, (ii) per-transaction traces via \texttt{debug\_traceTransaction}, and
(iii) on-chain bytecode and event logs for contracts interacting with known
builder clusters.

To identify builder-owned contracts, we combine:
\begin{packeditemize}
    \item the official whitelist of builder endpoints exposed through BSC’s PBS
          interface,
    \item address clusters with validator-operated endpoints,
    \item static analysis over contract creation patterns that repeatedly appear
          in builder-submitted blocks.
\end{packeditemize}

The resulting mapping yields 32{,}560 contract addresses linked to \bd{48Club}
and 30{,}781 to \bd{Blockrazor}. This labelling is conservative: it only
includes addresses whose behavioral signature (factory interactions, router
patterns, swap ordering templates) matches the builder’s known execution
pipeline.

\subsection{Active Builders and Their Shares}
\label{subsec-builder}

We first investigate all active builders. BSC's builder ecosystem comprises
nine identifiable brands (\bd{48Club}, \bd{Blockrazor}, \bd{Jetbldr},
\bd{Bloxroute}, \bd{Nodereal}, \bd{Blocksmith}, \bd{inblock}, \bd{Blockbus}, \bd{Flashblock}) operating a total of over
35 builder instances. We summarize them with their shares in
Table~\ref{tab:builder} in Appendix~\ref{apd-builderlist}.

\begin{figure}
    \centering
    \includegraphics[width=0.9\linewidth]{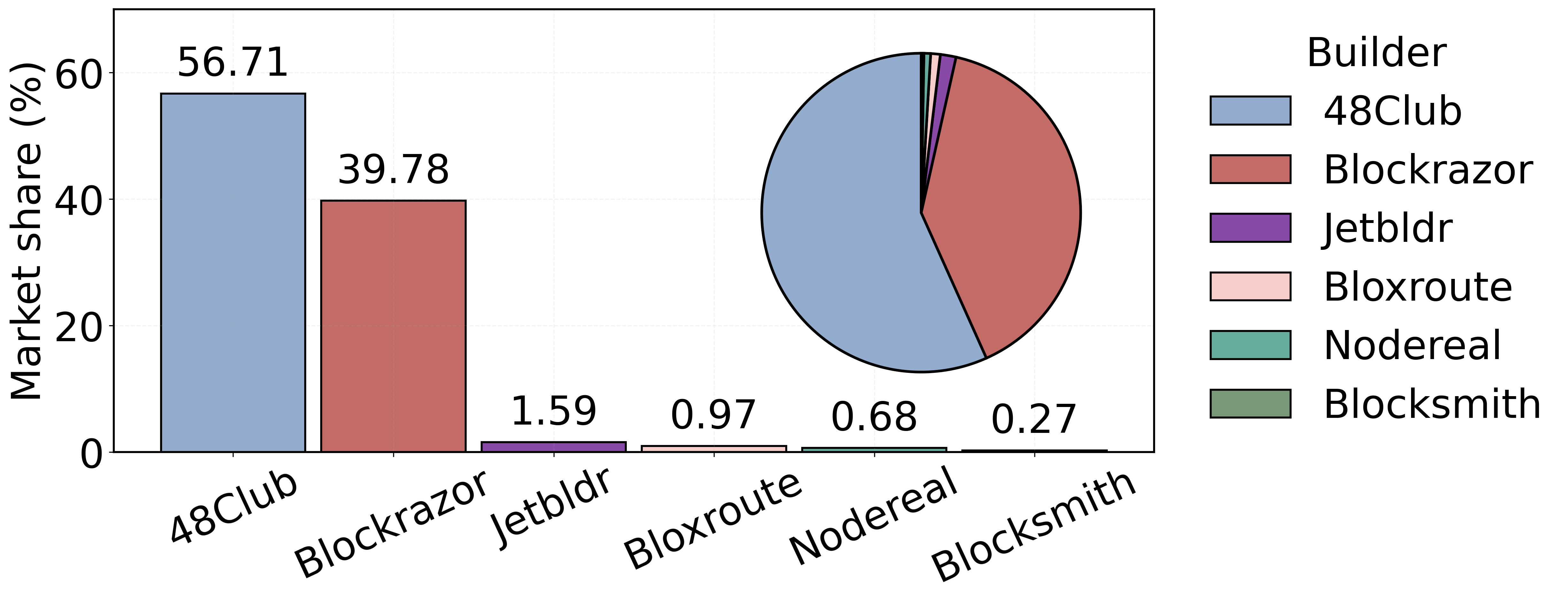}
    \caption{Market share distribution of Binance builders}
    \Description{A market-share chart shows Binance builders by block-construction share. 48Club is the largest builder, Blockrazor is second and close behind, while the remaining builders each hold only a very small share.}
    \label{fig:builder-share}
\end{figure}

We find that the market is highly concentrated (Fig.\ref{fig:builder-share}): \bd{48Club} leads with 17{,}965{,}461 blocks across 45 validators, controlling 46.57\% of builder share. \bd{Blockrazor} follows closely with 15{,}916{,}288 blocks (41.26\%, 45 validators). \bd{Jetbldr} ranks third but much smaller, producing 991{,}850 blocks (2.57\%, 40 validators). Low-tier participants include \bd{Bloxroute} (719{,}508 blocks, 1.86\%, 44 validators), \bd{Nodereal}
(383{,}758 blocks, 0.99\%, 36 validators) and \bd{Blocksmith} (147{,}228 blocks,
0.38\%, 35 validators). The other three builders remain a minor share.
The top two brands (\bd{48Club}, \bd{Blockrazor}) control over 87\% of all blocks. The remaining seven account for less than 13\% of blocks and an even smaller share of net arbitrage profit. For these minor builders, per-block profits are often indistinguishable from noise, and proposer payouts are sparse. 
Therefore, we conduct the measurements on these two dominant builders.

\subsection{Building Our Dataset}
\label{subsec-dataset}

We empirically collect on-chain activities from two builder entities on BSC.
Our observation window spans Apr.~1,~2025 to Feb.~10,~2026. Within this
window, builder-operated arbitrage becomes visible only after the deployment of
the relevant contracts:

\begin{packeditemize}
    \item \bd{48Club}, covering block ranges 50{,}636{,}154--80{,}336{,}045 and extending from late May~2025 to Feb.~10,~2026;
    
    \item \bd{Blockrazor}, ranging 51{,}651{,}126--78{,}675{,}307 and extending from mid-June~2025 to Feb.~01,~2026; its reported profit series is consolidated across a six-address operating cluster.
\end{packeditemize}

Our dataset is \texttt{arbitrage\_informations.csv}, available at
\textcolor{teal}{\url{https://anonymous.4open.science/r/MEV_Binance-5B2C}.}

\vspace{3pt}
\noindent\textbf{Why do the builder traces begin in May 2025?}
Before late May~2025, most observed arbitrage on BSC was still executed by independent searchers, while builders mainly acted as intermediaries and took a share of submitted bundles. From late May onward, we observe for the first time that builders themselves deploy large-scale arbitrage contracts and directly capture opportunities that previously belonged to searchers. We do not claim that this is the absolute starting point of such activity, nor that our dataset is fully exhaustive, but it is the earliest large-scale builder-operated deployment that we can verify on-chain. This timing is also consistent with community reports~\cite{everything,bnbdiscuss1,bnbdiscuss2}.

\vspace{3pt}
\noindent\textbf{What the data look like?}
Our raw trace dataset covers 9.63 M labeled builder-executed arbitrage cycles. Major stablecoins and wrapped
assets account for about \textbf{66.7\%} of total routing volume
(Fig.~\ref{fig:mev-asset-volume}); within these, \texttt{USDT}, \texttt{USD1},
\texttt{USDC}, and \texttt{WBNB} dominate path usage. This motivates focusing
subsequent analysis on these four bases.

\subsection{Arbitrage Identification}
\label{subsec:arbitrage}

For each transaction initiated by a builder-owned contract, we reconstruct
the executed swap path by parsing \texttt{Swap}, \texttt{Sync}, and router
events. A transaction is labeled as an arbitrage cycle if its entry and exit
assets match, i.e., $\mathsf{asset}_{\mathrm{in}} = \mathsf{asset}_{\mathrm{out}}$.
Multi-hop paths (e.g., USDT $\rightarrow$ WBNB $\rightarrow$ USD1 $\rightarrow$
USDT) are treated as a single atomic opportunity.
For each cycle, we compute:
\begin{packeditemize}
    \item \textbf{Gross profit.} difference between input and final output
          (in token units);
    \item \textbf{Share-profit.} amount redirected to designated redistribution
          addresses (e.g., \texttt{0xffff\ldots fffe});
    \item \textbf{Net profit.} gross minus share-profit and gas fees, converted
          into BNB.
\end{packeditemize}

These fields allow us to distinguish builder-internal redistribution from
searcher-facing MEV payouts. A distinctive feature of BSC’s PBS is the presence
of \textit{share profit}, where part of the arbitrage gains is diverted to
special addresses such as \texttt{0xffff\ldots fffe}. In our traces, this
address acts as a validator-revenue sink rather than the official BEP-322
payout contract; the formal payout path is implemented through a separate
zero-gas accounting transaction (Appendix~\ref{apd-method}).
In addition, we observe cases where profits are routed directly into liquidity
pools (e.g., the \texttt{USDT}--\texttt{WBNB} pair at
\texttt{0x16b9\ldots 0daE}).

%=================================================
\section{Empirical Results}
\label{sec-result}
%=================================================

We now present our empirical results. We provide a view of \textit{who} captures MEV on BSC (\S\ref{subsec-profBuild}, \S\ref{subsec:proValid}), \textit{how} these profits are made (token choice in \S\ref{subsec-profToken} and swap path in \S\ref{subsec:swap-complexity}, \S\ref{subsec:path-efficiency}),  \textit{whether} the system is trending toward greater inequality (\S\ref{subsec:mann-kendall}) and \textit{where} extraction is concentrated (\S\ref{subsec:centralisation-risk}, \S\ref{subsec:asset-flow}).

\begin{figure}[!htb]
    \centering
    \subfigure[Profit spread]{
    \begin{minipage}{0.46\linewidth}
    \centering
    \includegraphics[width=\linewidth]{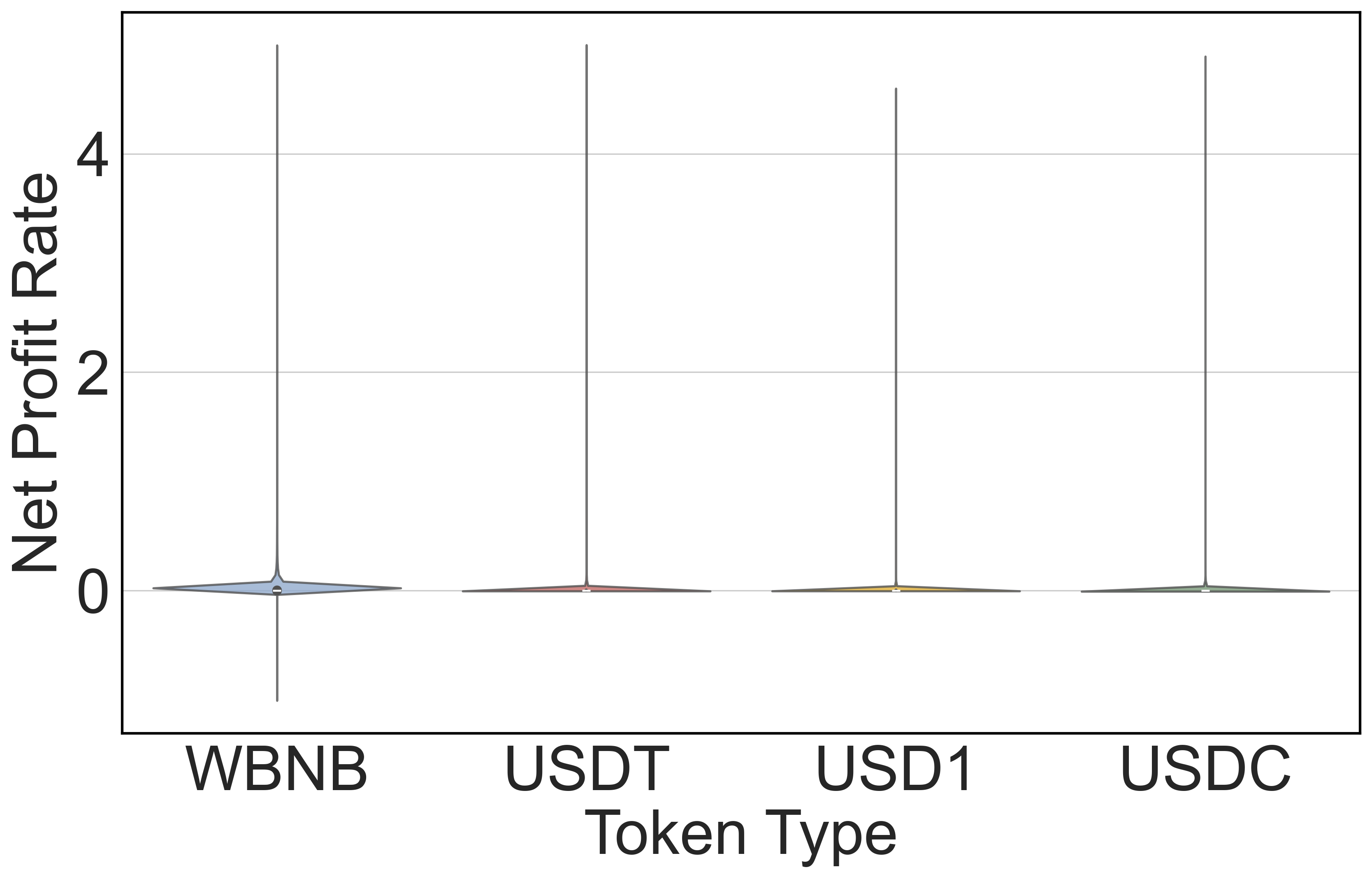}
    \end{minipage}
    \label{fig:anova_violin}
    }
    \subfigure[Distribution shape]{
    \begin{minipage}{0.46\linewidth}
    \centering
    \includegraphics[width=\linewidth]{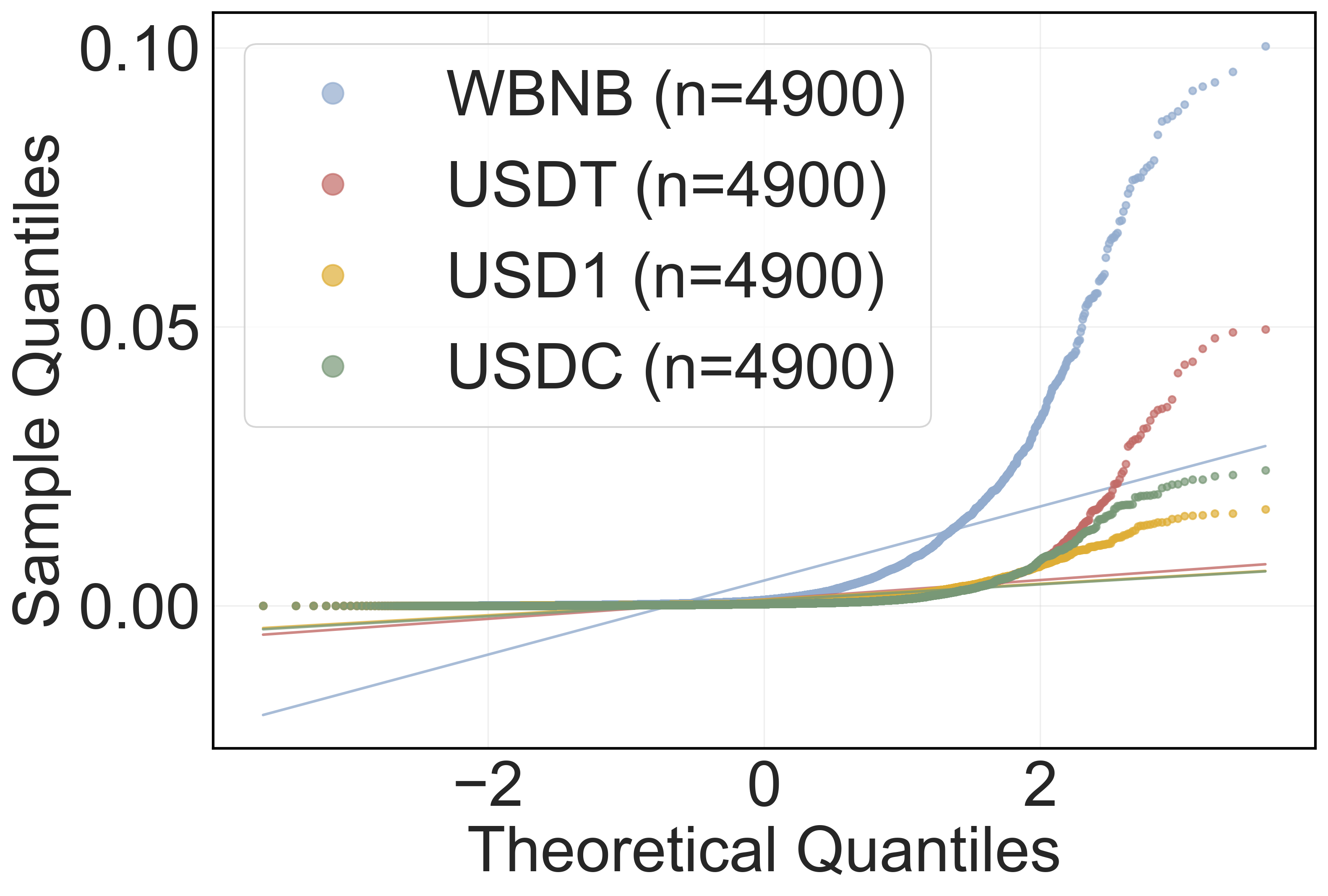}
    \end{minipage}
    \label{fig:anova_qq}
    }
    \caption{Profitability patterns across token types.}
    \Description{Two panels summarize profit-rate distributions across token types. A violin plot shows that most trades cluster near zero profit, while a Q-Q plot shows the heaviest positive tails for WBNB and USDT compared with USD1 and USDC.}
    \label{fig:anova_combined}
\end{figure}

% --------------------------------------
\subsection{Profits by Token (Fig.\ref{fig:anova_combined})}
\label{subsec-profToken}
% --------------------------------------

We examine whether the base token affects MEV builder profitability on BSC, focusing on four major tokens.

Most trades across all tokens (Fig.\ref{fig:anova_violin}) still yield very small net profit after gas costs, with densities tightly concentrated around a profit rate close to zero. This supports the view that BSC’s arbitrage market is strongly competed, where most opportunities are quickly arbitraged away and only basis-point spreads remain available.

The tails, however, differ by token. In the Q–Q plots (Fig.\ref{fig:anova_qq}), \texttt{WBNB} shows the heaviest right tail, with a small set of trades pushing sample quantiles well above the normal line, followed by a milder but visible tail for \texttt{USDT}. By contrast, \texttt{USD1} and \texttt{USDC} remain more tightly clustered around the diagonal, with fewer and smaller deviations at the upper end. This indicates that while all four tokens exhibit non-Gaussian, skewed payoffs, outsized wins are more likely on \texttt{WBNB} and \texttt{USDT} routes than on \texttt{USD1}/\texttt{USDC}.

We conclude that MEV on BSC is predominantly low-margin, but a handful of high-return outliers, especially in \texttt{WBNB} and \texttt{USDT} paths, contribute disproportionately to aggregate builder revenue.

\begin{figure}[!htbp]
    \centering
    % Absolute profit
    \subfigure[Absolute profit]{%
    \begin{minipage}{0.45\linewidth}
    \centering
    \includegraphics[width=\linewidth]{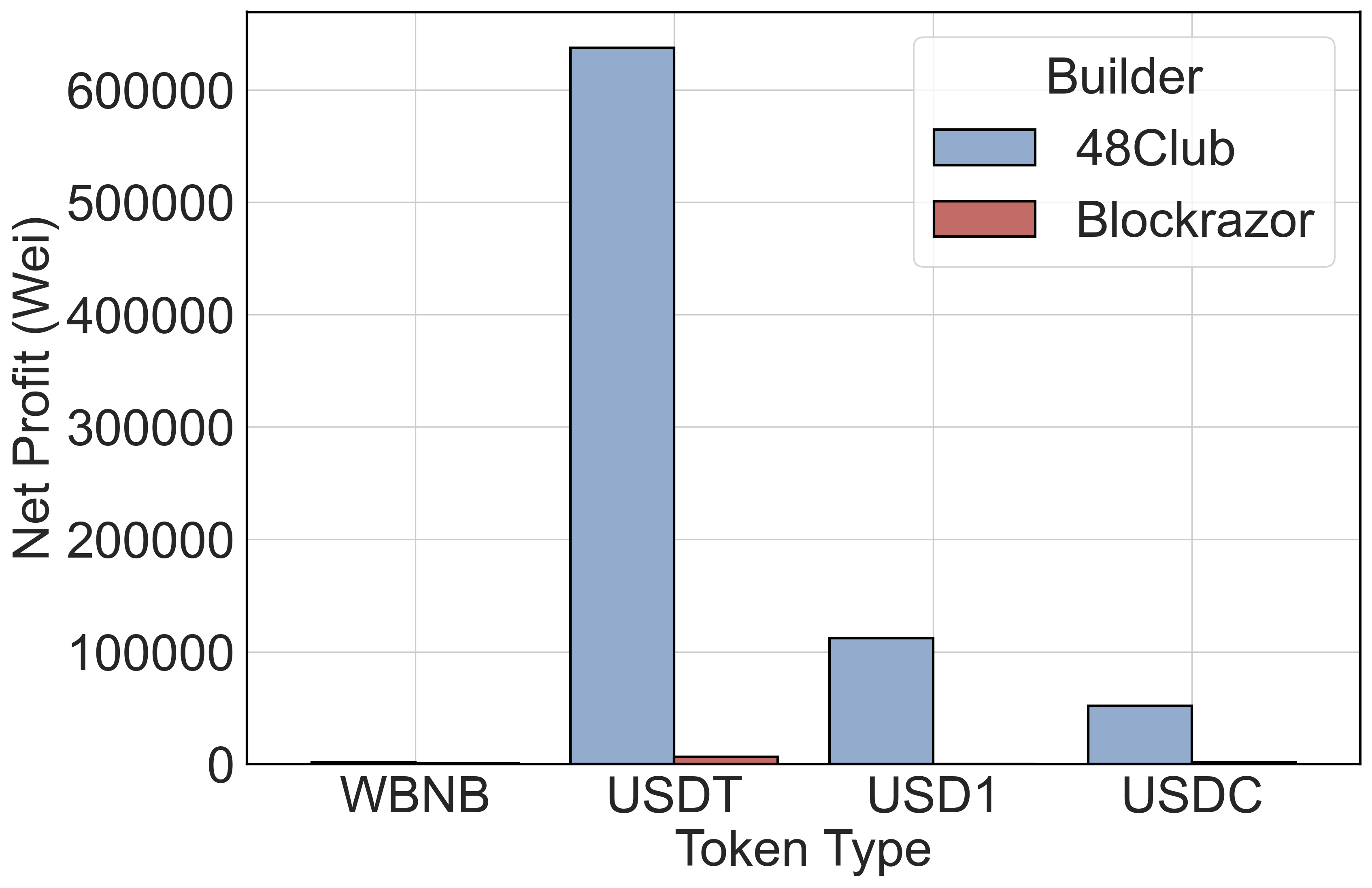}
    \label{fig:chisquare_absolute}
    \end{minipage}
    }
    % Contingency heatmap
    \subfigure[Hotspots]{%
    \begin{minipage}{0.45\linewidth}
    \centering
    \includegraphics[width=\linewidth]{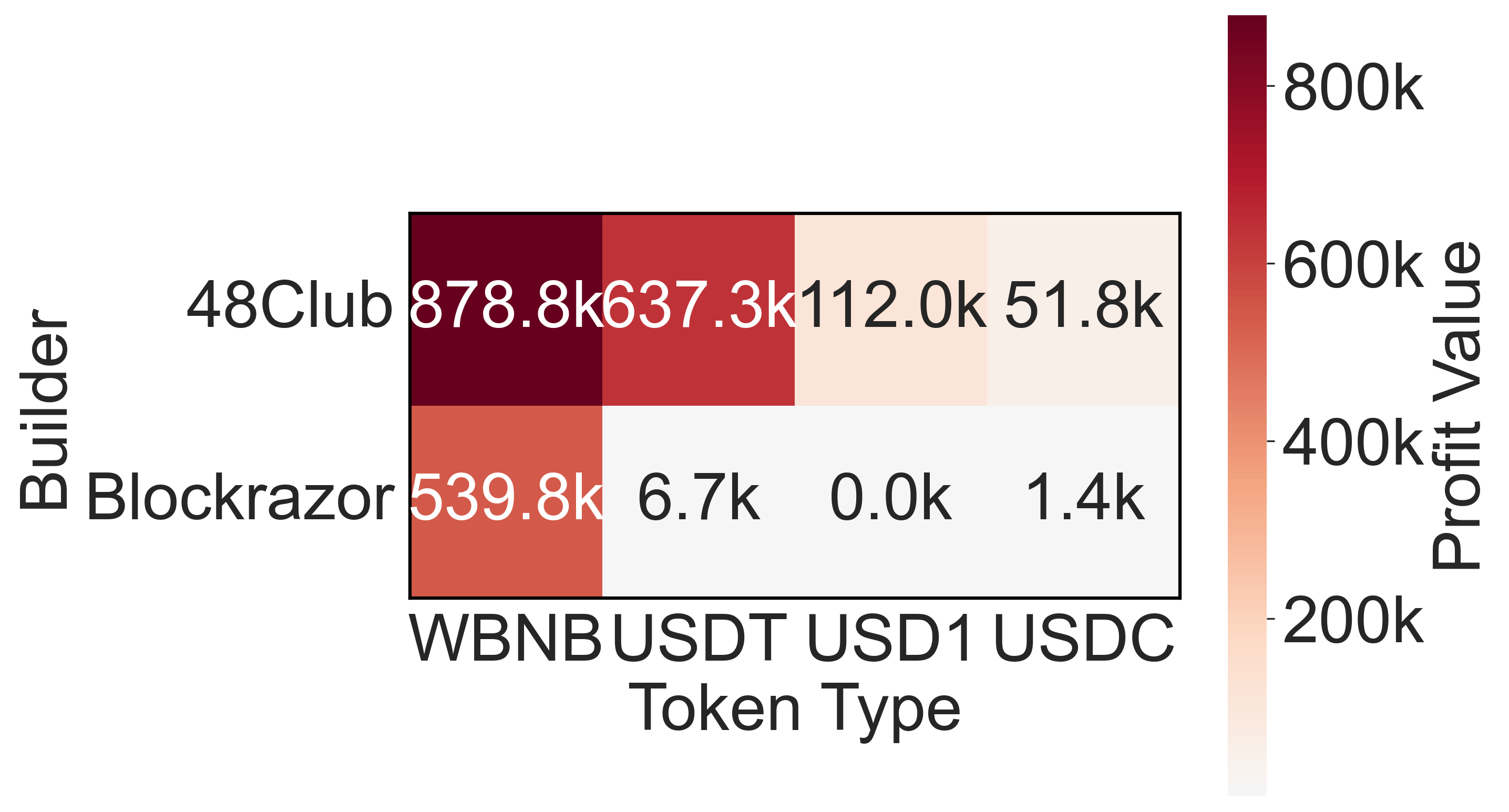}
    \label{fig:chisquare_heatmap}
    \end{minipage}
    }
    
    % Percentage profit
    \subfigure[Market share]{%
    \begin{minipage}{0.45\linewidth}
    \centering
    \includegraphics[width=\linewidth]{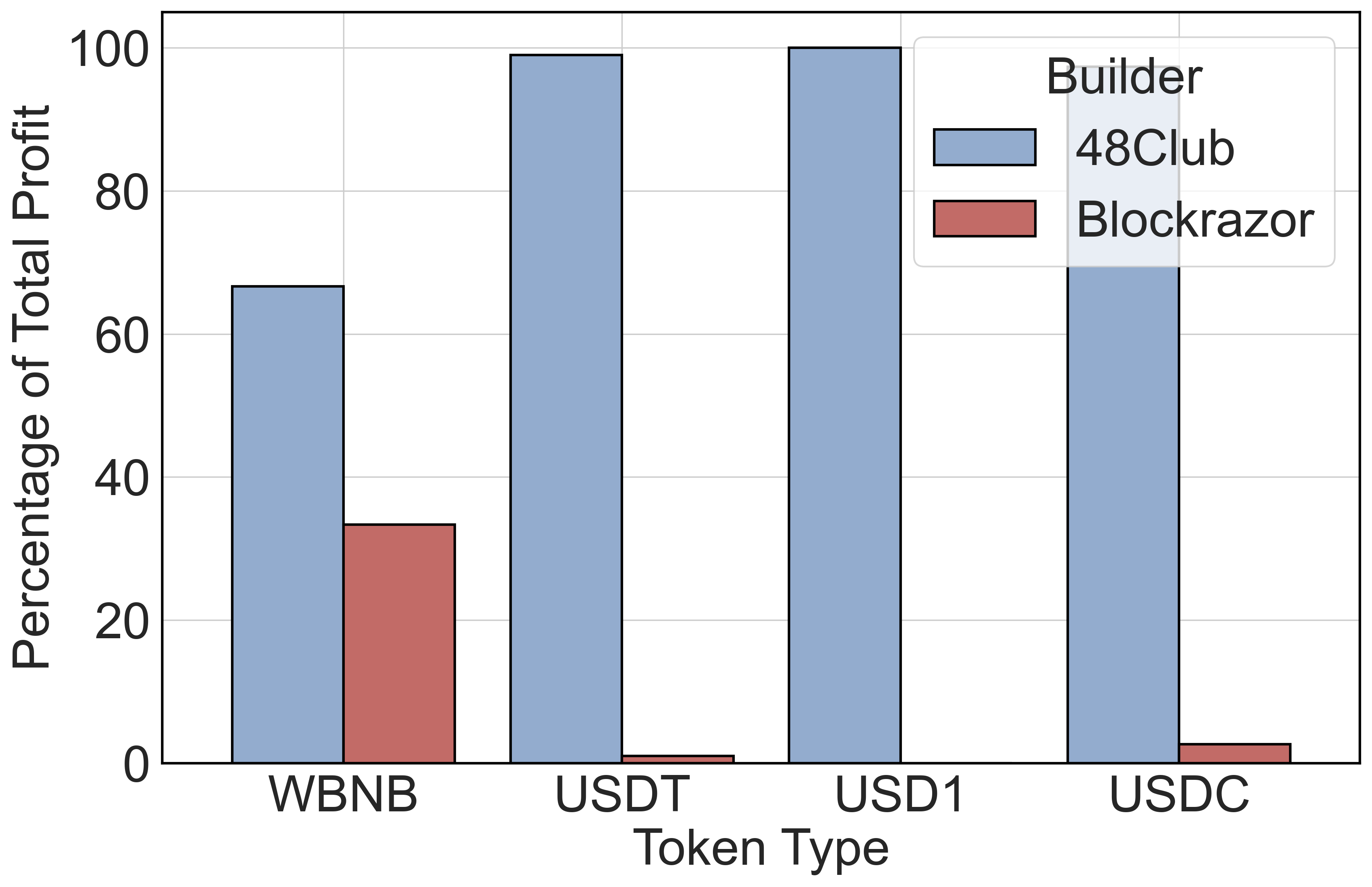}
    \label{fig:chisquare_percentage}
    \end{minipage}
    }
    % Log-scaled profit
    \subfigure[Log scale]{%
    \begin{minipage}{0.45\linewidth}
    \centering
    \includegraphics[width=\linewidth]{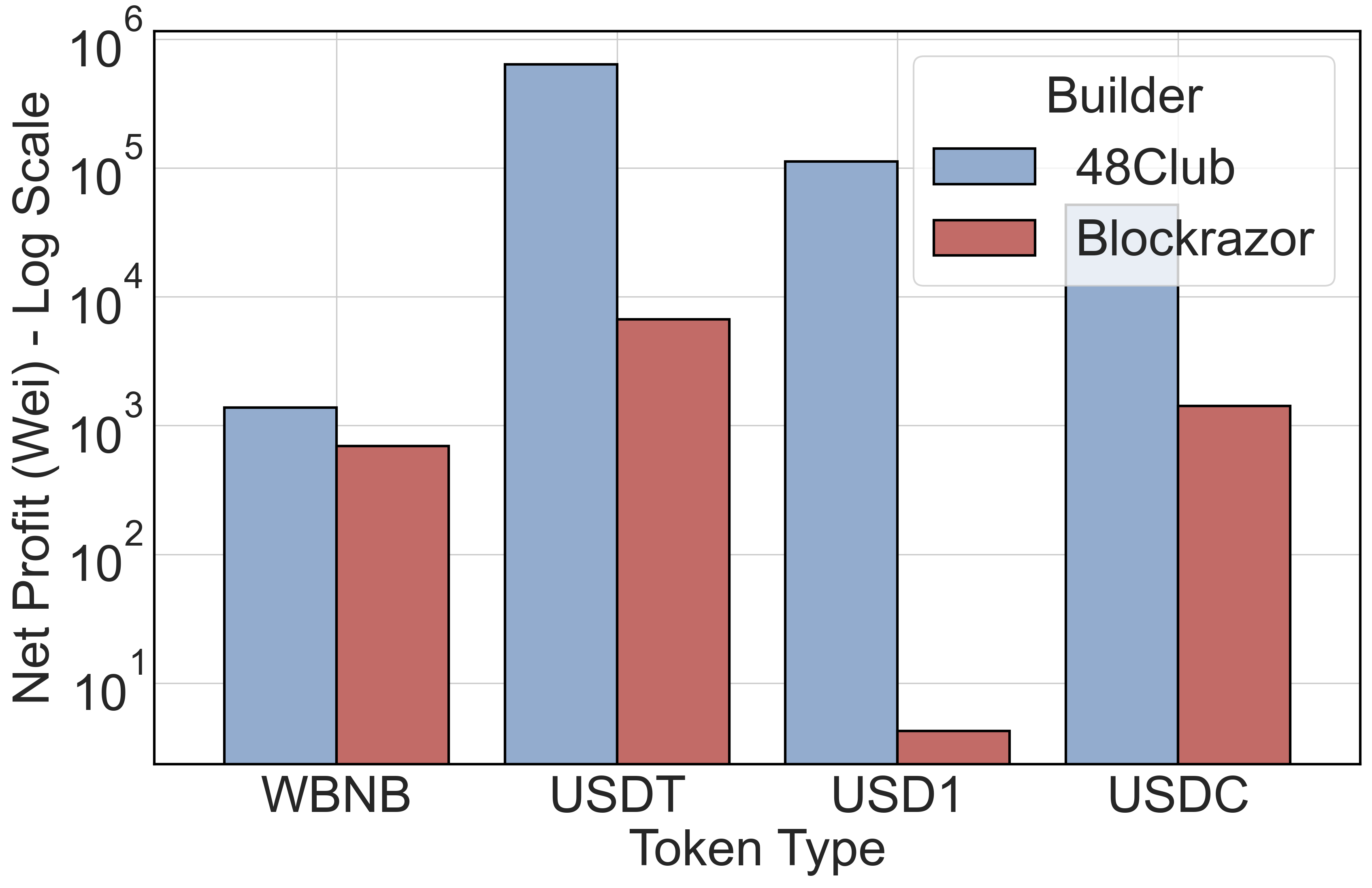}
    \label{fig:chisquare_log}
    \end{minipage}
    }

    \caption{MEV profit concentration by builder and token. 48Club dominates overall, while Blockrazor remains material mainly in \texttt{WBNB}.} 
    \label{fig:chisquare_combined}
\end{figure}

% --------------------------------------
\subsection{Profits by Builder (Fig.\ref{fig:chisquare_combined})}
\label{subsec-profBuild}
% --------------------------------------

For consistent builder comparisons, we convert all profits to \texttt{USD} using fixed rates: \texttt{WBNB}=$635.95$ and $\texttt{USDT}/\texttt{USD1}/\texttt{USDC}=1$. A daily WBNB/USD series slightly changes absolute totals but preserves all rankings and concentration patterns; we therefore report fixed-rate values for readability.

The results show a strong imbalance (Fig.\ref{fig:chisquare_absolute}). \bd{48Club} dominates overall, retaining roughly 1.68\,M USD in total net profit, while \bd{Blockrazor} retains about 0.55\,M USD. For \bd{48Club}, \texttt{WBNB} is the largest contributor (about 0.88\,M USD), followed by \texttt{USDT} (about 0.64\,M USD), \texttt{USD1} (about 0.11\,M USD), and a smaller tail in \texttt{USDC} (about 0.05\,M USD). By contrast, \bd{Blockrazor}'s activity is concentrated in \texttt{WBNB} (about 0.54\,M USD), with only a very small tail in the stablecoins.

Per-token shares (Fig.\ref{fig:chisquare_percentage}) further underline this concentration. In \texttt{WBNB}, \bd{48Club} captures roughly two thirds of total profit, while \bd{Blockrazor} takes the remaining one third. In the three stablecoins, however, \bd{48Club} is close to a monopoly: it accounts for about 99\% of profit in \texttt{USDT}, essentially all profit in \texttt{USD1}, and more than 97\% in \texttt{USDC}. The heatmap in Fig.\ref{fig:chisquare_heatmap} highlights the \bd{48Club}--\texttt{WBNB} cell as the primary hotspot, with \bd{48Club}'s \texttt{USDT} position as the second-largest profit center.

On a log scale (Fig.~\ref{fig:chisquare_log}), \bd{Blockrazor} remains visible for \texttt{WBNB} but drops sharply for stablecoins. For \bd{48Club}, profits follow \texttt{WBNB} $>$ \texttt{USDT} $>$ \texttt{USD1} $>$ \texttt{USDC}. For \bd{Blockrazor}, profits are almost entirely \texttt{WBNB}-driven, with negligible stablecoin gains. Overall, BSC MEV profit is concentrated by both builder and token: \bd{48Club} captures most extracted value, while \bd{Blockrazor} competes meaningfully only in \texttt{WBNB}.

\begin{figure}[!htbp]
    \centering
    \subfigure[Profit split by token]{%
    \begin{minipage}{0.45\linewidth}
    \centering
    \includegraphics[width=\linewidth]{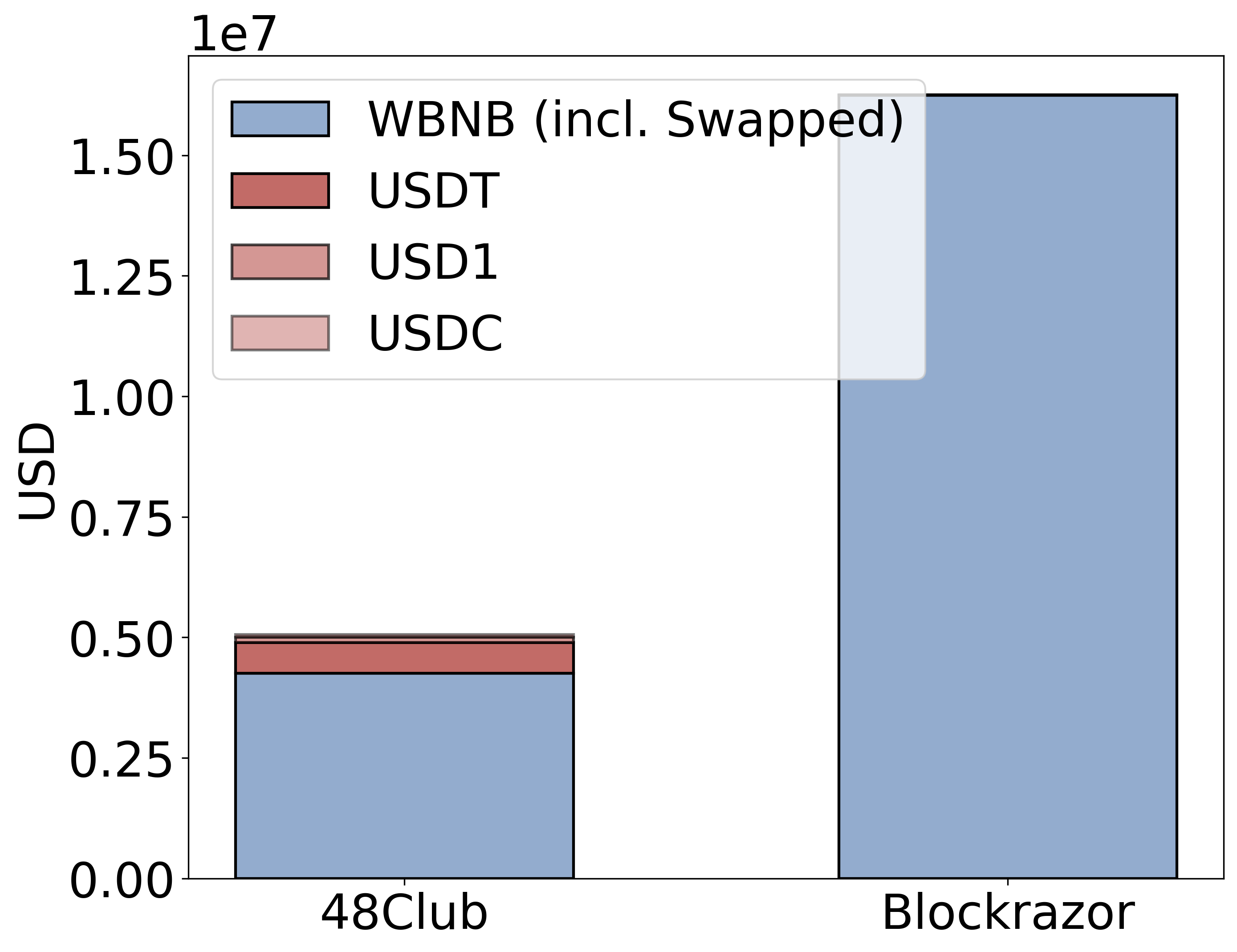}
    \label{fig:chart1_validator_share_by_token}
    \end{minipage}
    }
    \subfigure[\texttt{WBNB} vs swapped flows]{%
    \begin{minipage}{0.45\linewidth}
    \centering
    \includegraphics[width=\linewidth]{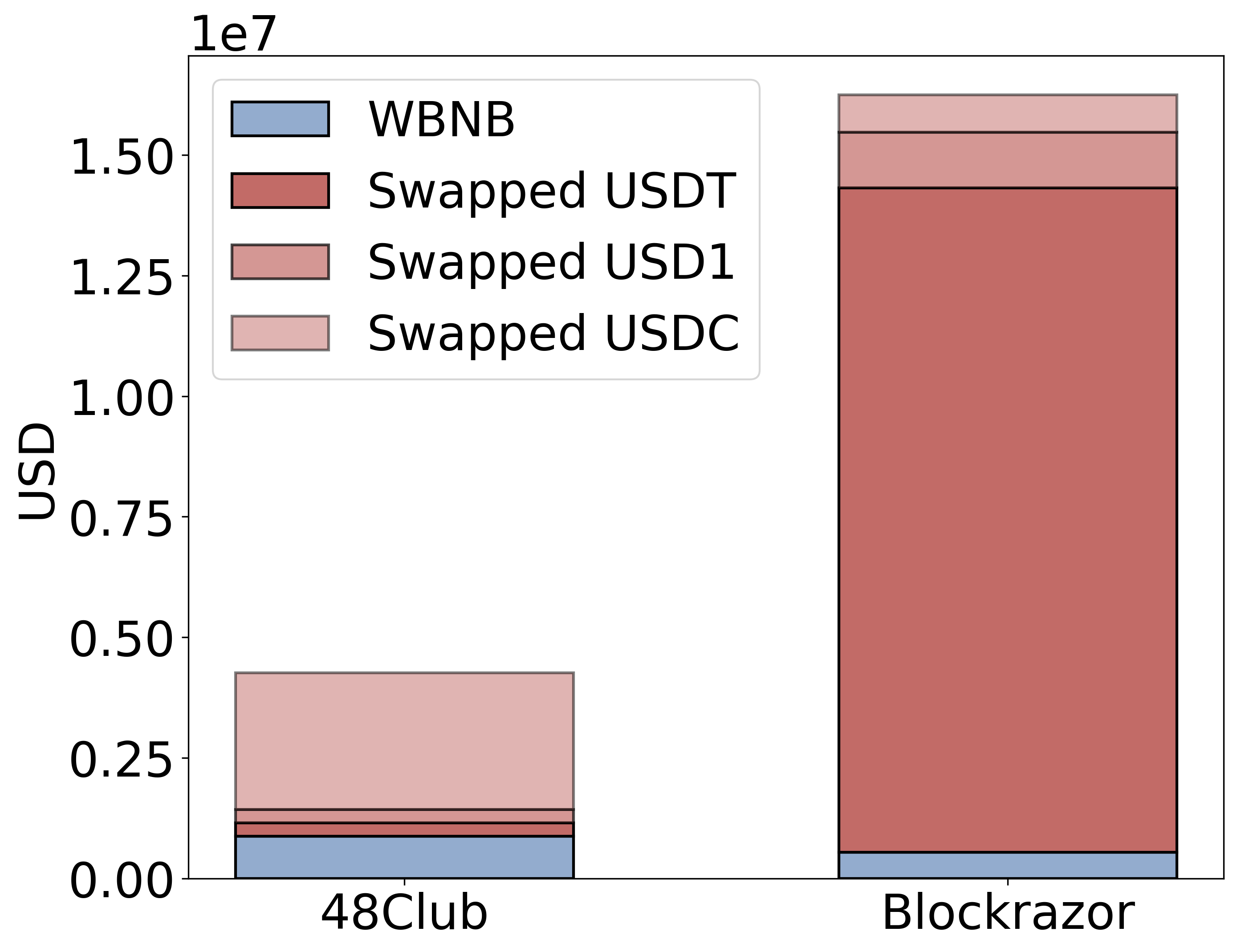}
    \label{fig:chart2_wbnb_breakdown}
    \end{minipage}
    }
    
    \subfigure[\bd{48Club} net vs proposer payouts]{%
    \begin{minipage}{0.45\linewidth}
    \centering
    \includegraphics[width=\linewidth]{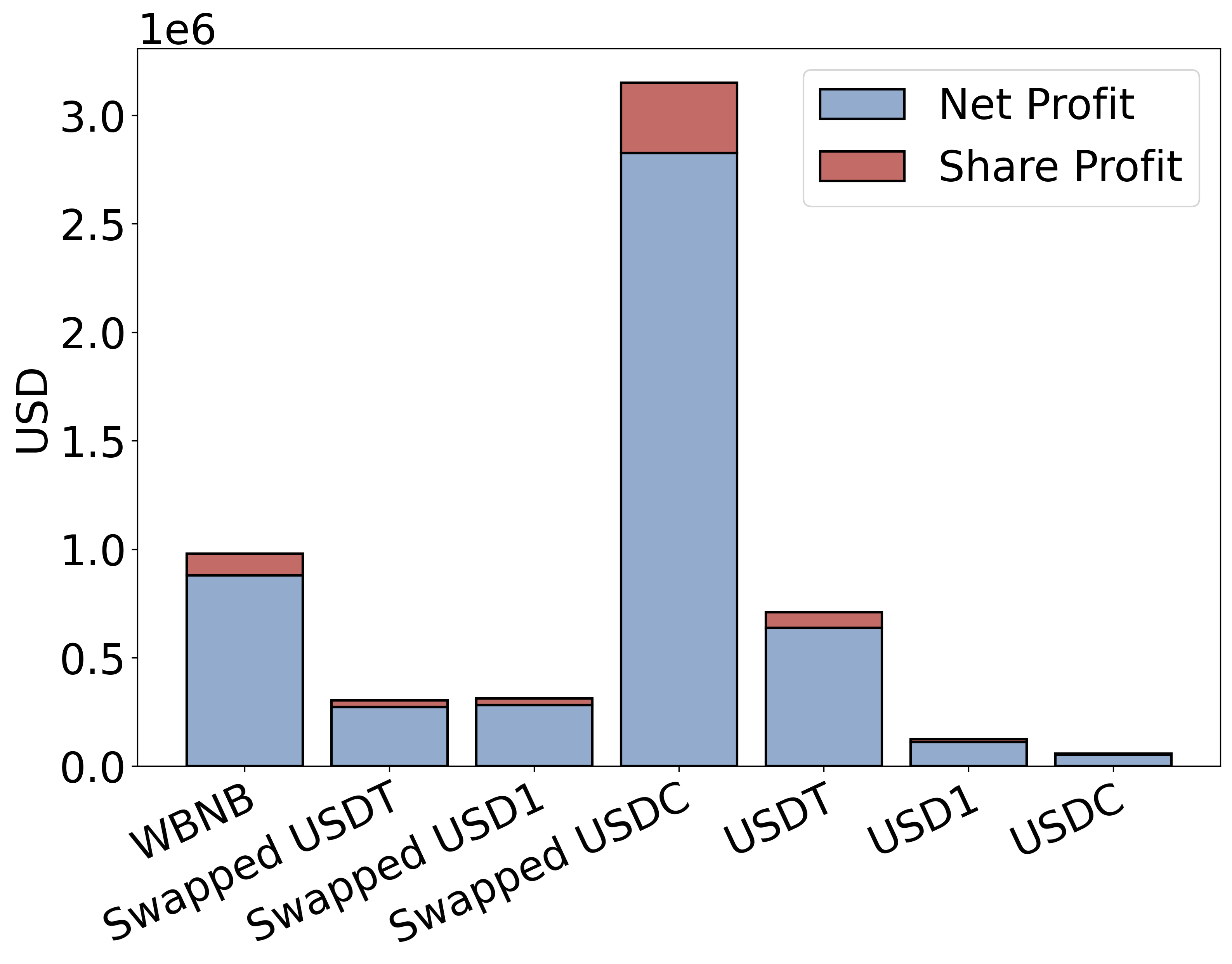}
    \label{fig:cchart3_48club_net_share}
    \end{minipage}
    }
    \subfigure[\bd{Blockrazor} net vs proposer payouts]{%
    \begin{minipage}{0.45\linewidth}
    \centering
    \includegraphics[width=\linewidth]{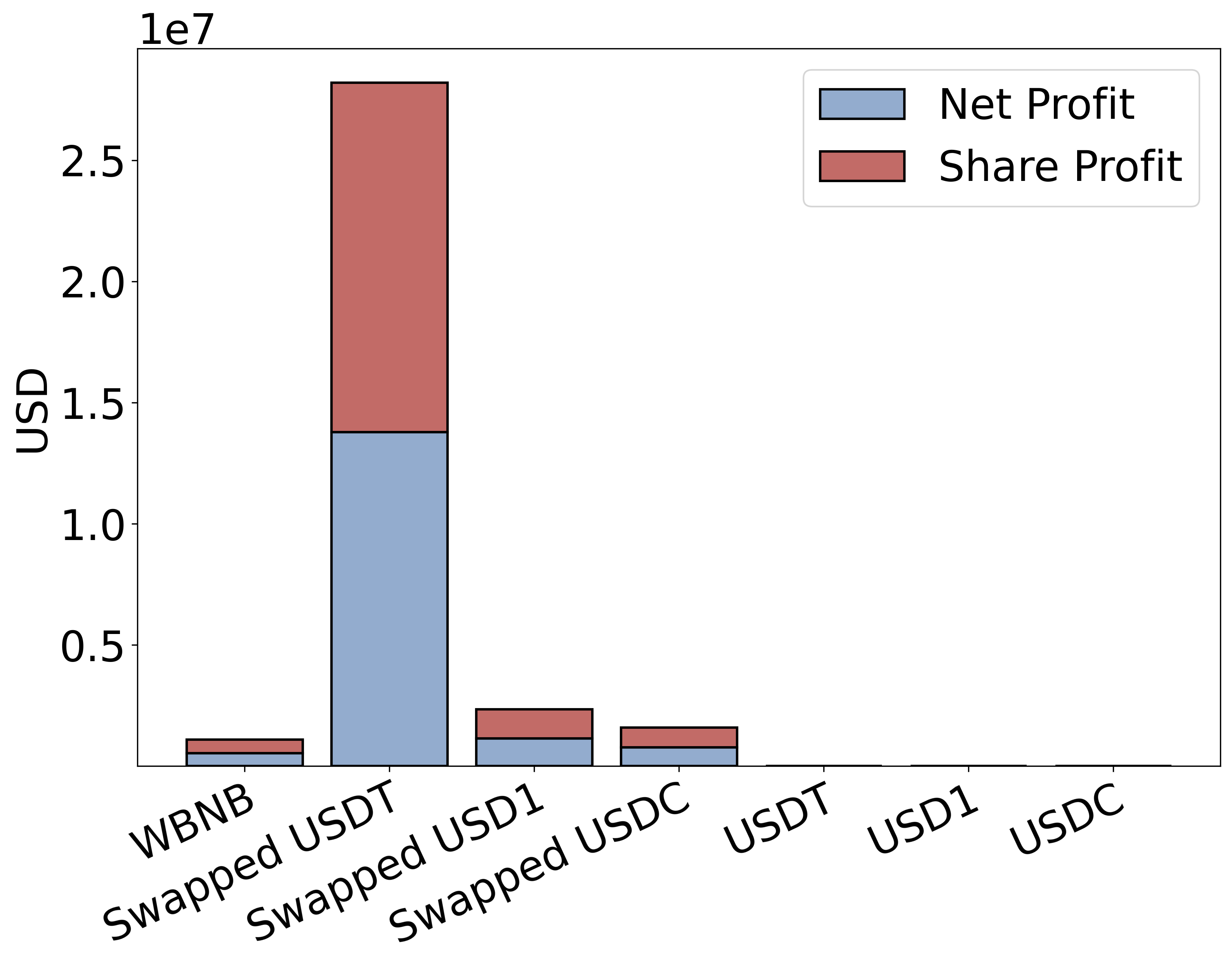}
    \label{fig:chart4_blockrazor_net_share}
    \end{minipage}
    }

    \caption{Proposer payouts by builder. 48Club spreads payouts across tokens; Blockrazor concentrates on swapped \texttt{USDT}.}
    \label{fig:validator_profit_combined}
\end{figure}

% --------------------------------------
\subsection{Profits by Proposer (Fig.\ref{fig:validator_profit_combined})}
\label{subsec:proValid}
% --------------------------------------

We also examine how profits are split between builders and proposers. 
Fig.\ref{fig:chart1_validator_share_by_token} shows that \bd{48Club} distributes proposer payouts across tokens, with \texttt{WBNB} (including swapped flows) as the dominant contributor and \texttt{USDT} as a clear secondary component, followed by smaller shares from \texttt{USD1} and \texttt{USDC}. In contrast, \bd{Blockrazor}'s proposer payouts are dominated by \texttt{WBNB}, while others remain minor.

Decomposing \texttt{WBNB} reveals structural differences. For \bd{48Club}, proposer payouts are still anchored by native \texttt{WBNB}, but they are supplemented by swapped stablecoins, especially Swapped \texttt{USDT}, with smaller additions from Swapped \texttt{USD1} and Swapped \texttt{USDC}. For \bd{Blockrazor}, proposer payouts are overwhelmingly driven by Swapped \texttt{USDT}, with only a thin native \texttt{WBNB} layer and smaller swapped \texttt{USD1}/\texttt{USDC} components. This shows that \bd{Blockrazor}'s proposer incentives are primarily derived from a narrow stablecoin-heavy arbitrage flow rather than direct \texttt{WBNB} profits.

The results suggest two different styles: \bd{48Club} spreads activities across multiple tokens and keeps proposer compensation diversified, while \bd{Blockrazor} channels proposer payouts through a much narrower route set centered on swapped \texttt{USDT}. Importantly, the new figures no longer support the older shorthand that \bd{Blockrazor} simply ``pays validators less'': instead, \bd{Blockrazor}'s proposer payouts are large in absolute terms but highly concentrated, whereas \bd{48Club} spreads both retained profit and proposer payouts across more tokens. These patterns reinforce our interpretation that \bd{48Club} pursues broader proposer alignment, whereas \bd{Blockrazor} relies on a smaller number of concentrated arbitrage flows.

\begin{figure}[!htbp]
    \centering
    \subfigure[Swap counts]{%
        \begin{minipage}{0.45\linewidth}
        \centering
        \includegraphics[width=\linewidth]{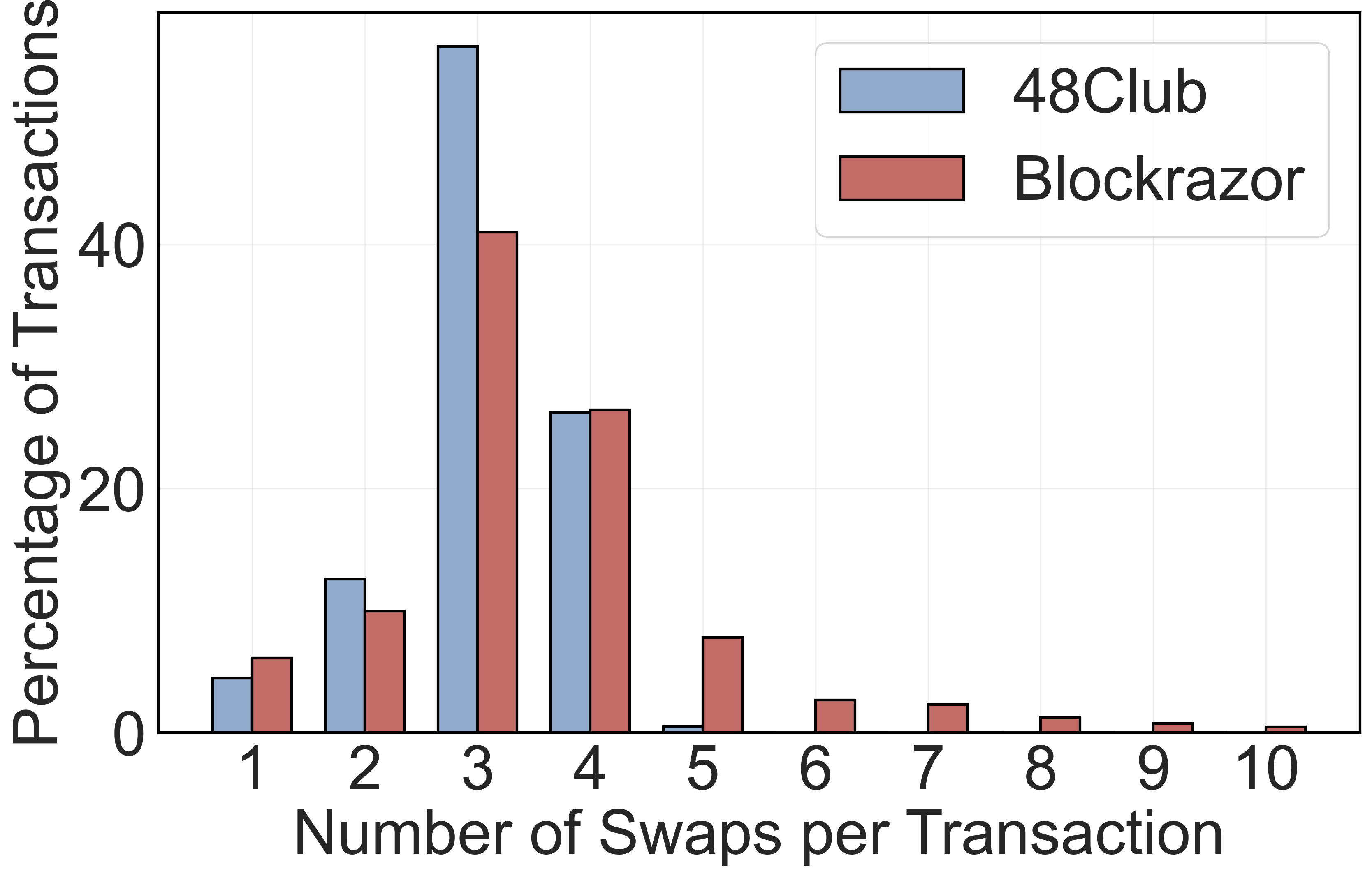}
        \label{fig:ks_freq}
        \end{minipage}
    }
    \hfill
    \subfigure[Cumulative distribution]{%
        \begin{minipage}{0.45\linewidth}
        \centering
        \includegraphics[width=\linewidth]{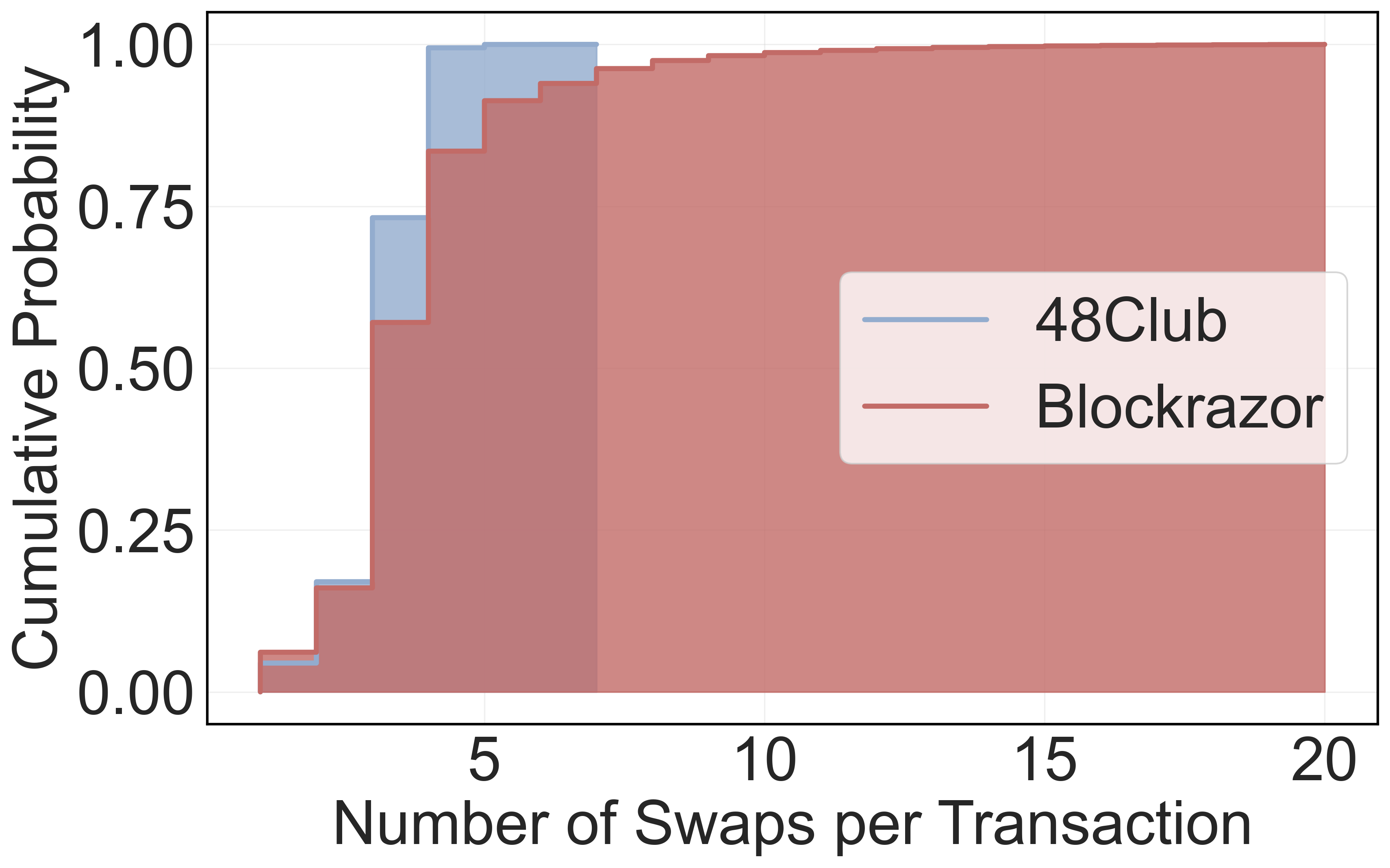}
        \label{fig:ks_ecdf}
        \end{minipage}
    }
    \caption{Swap-path complexity by builder. Most trades use short routes, while Blockrazor shows a longer tail.}
    \label{fig:ks_combined}
\end{figure}

% --------------------------------------
\subsection{Swap Paths by Builder (Fig.\ref{fig:ks_combined})}
\label{subsec:swap-complexity}
% --------------------------------------

A key question is how builders design their arbitrage routes. Fig.\ref{fig:ks_freq} and \ref{fig:ks_ecdf} show the distribution of swap counts per transaction for \bd{48Club} and \bd{Blockrazor}.

Most trades from both builders still rely on short paths: two- and three-swap routes account for the majority of their activity, confirming that simple arbitrage and triangular cycles remain the dominant strategies on BSC. However, the height of the bars in Fig.\ref{fig:ks_freq} reveals subtle differences. \bd{48Club} is heavily concentrated in two-swap bundles, with three-swap paths as a clear second choice, whereas \bd{Blockrazor} spreads more mass into three- and four-swap transactions and shows a gradual drop-off as path length increases.

The tail behavior (Fig.\ref{fig:ks_ecdf}) is clearer in the cumulative distribution. Almost all of \bd{48Club}’s transactions finish by four to five swaps, while \bd{Blockrazor} continues to take routes with up to 10--20 hops, even if these account for only a small share of its flow. This suggests that \bd{48Club} prioritizes speed and gas efficiency by concentrating on short paths, while \bd{Blockrazor} occasionally experiments with longer multi-hop strategies that may capture niche spreads at higher cost. Any hop limits or stricter gas-based penalties would therefore fall more heavily on \bd{Blockrazor}, underlining the need for mitigation measures that account for heterogeneous builder styles rather than assuming a single routing pattern.

\begin{figure}[!ht]
    \centering
    \subfigure[Correlation by builder]{%
    \begin{minipage}[t]{0.45\linewidth}
        \centering
        \includegraphics[width=\linewidth]{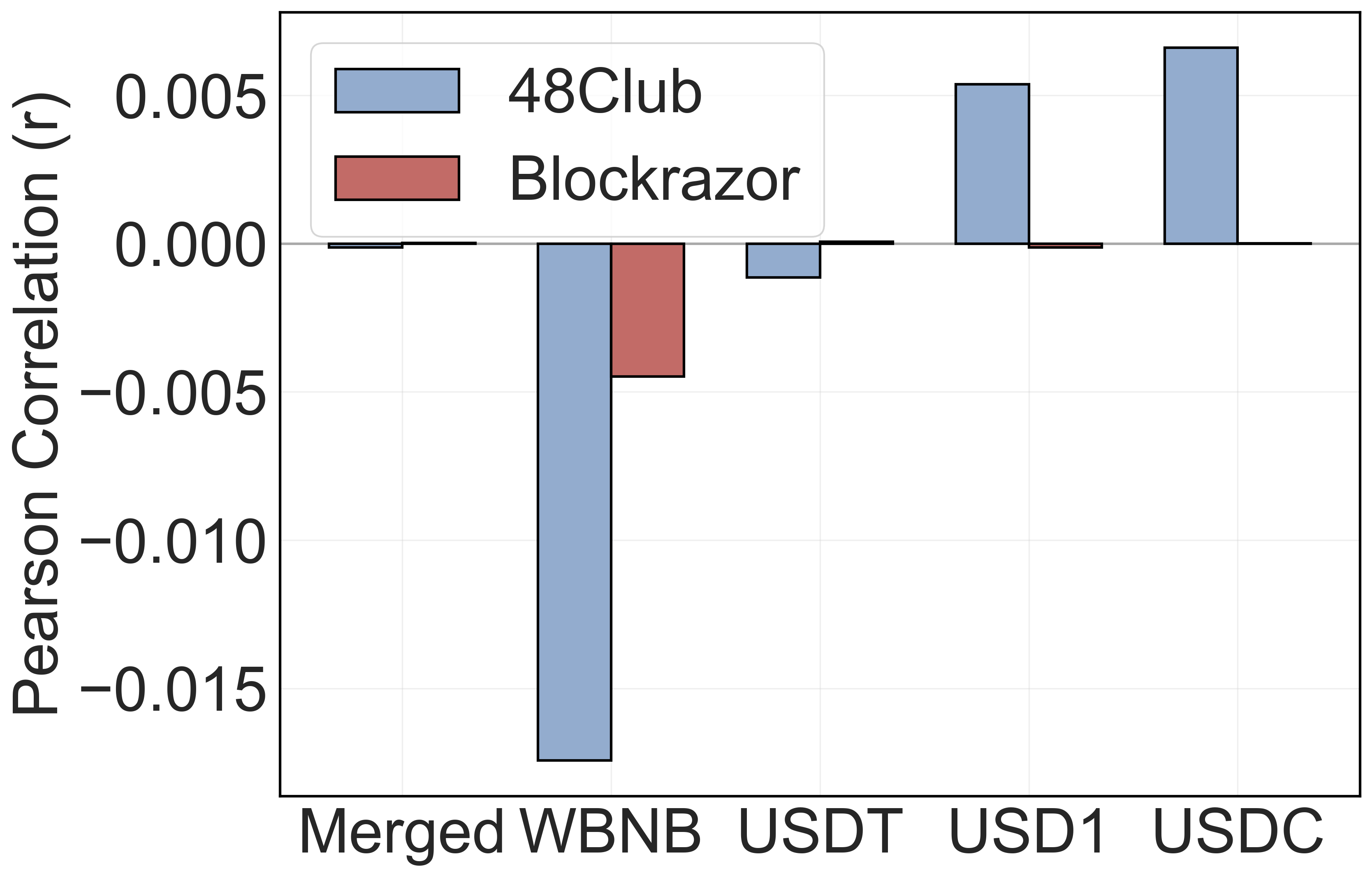}
        \label{fig:pearson-bar}
    \end{minipage}
    }
    \hfill
    \subfigure[Efficiency distribution]{%
    \begin{minipage}[t]{0.45\linewidth}
        \centering
        \includegraphics[width=\linewidth]{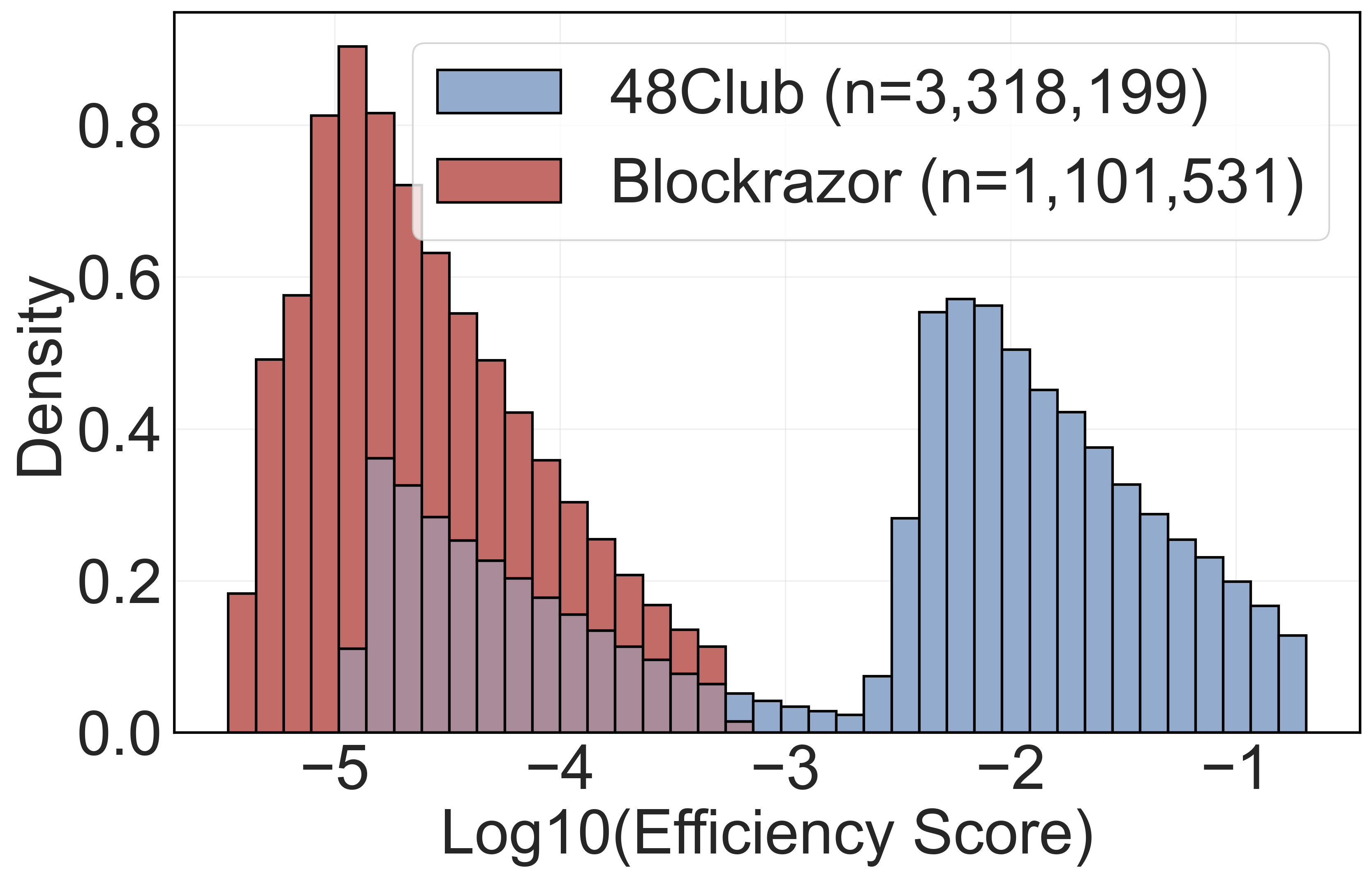}
        \label{fig:pearson-effdist}
    \end{minipage}
    }

    \subfigure[Correlation matrix]{%
    \begin{minipage}[t]{0.45\linewidth}
        \centering
        \includegraphics[width=\linewidth]{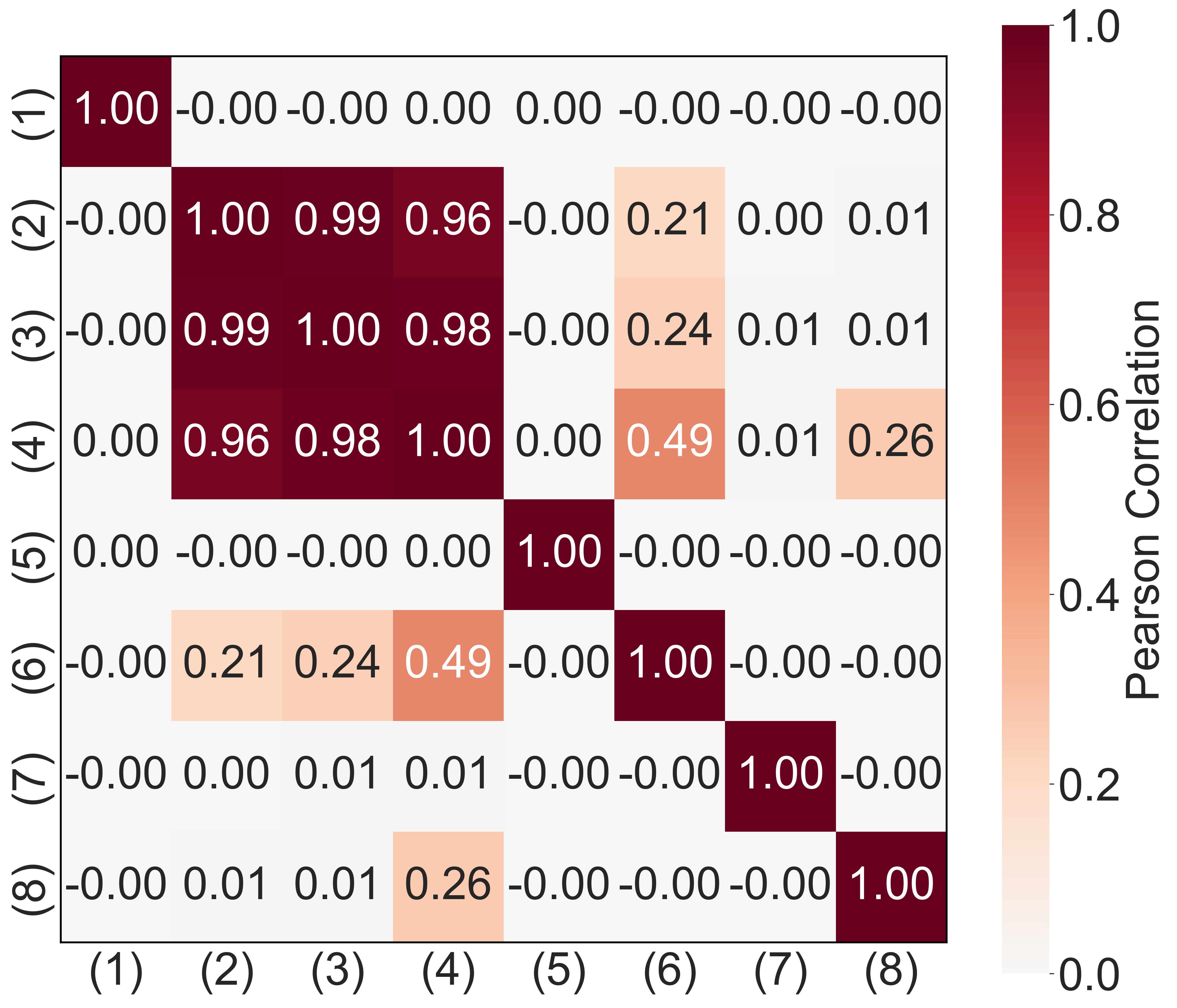}
        \label{fig:pearson-matrix}
    \end{minipage}
    }
    \hfill
    \subfigure[Swap count vs.\ profit]{%
    \begin{minipage}[t]{0.45\linewidth}
        \centering
        \includegraphics[width=\linewidth]{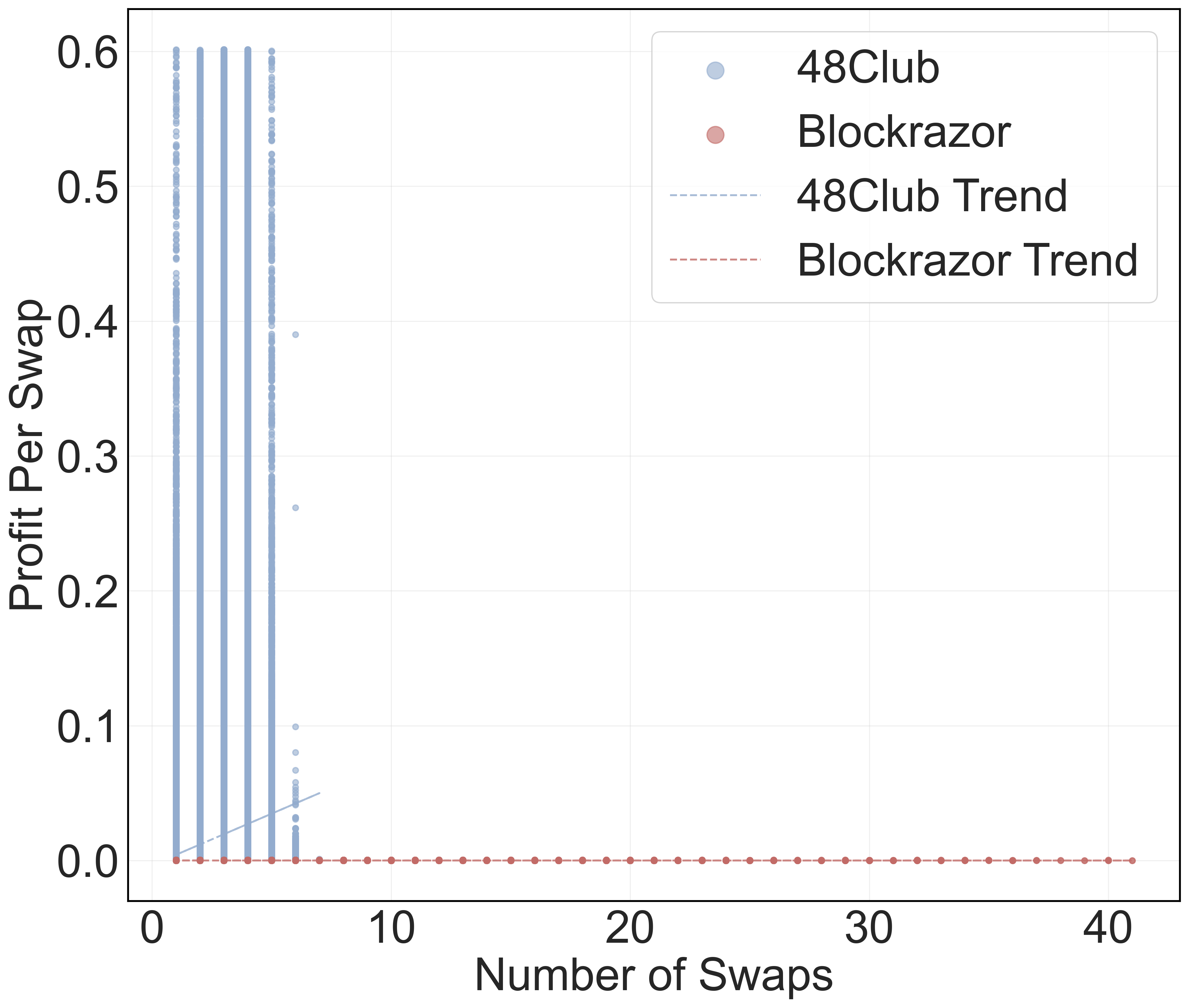}
        \label{fig:pearson-scatter}
    \end{minipage}
    }

    \caption{Builder efficiency by path length. Longer paths add little profit, and 48Club remains more efficient.}
    \label{fig:pearson-suite}
\end{figure}

% --------------------------------------
\subsection{Profits by Path Length (Fig.\ref{fig:pearson-suite})}
\label{subsec:path-efficiency}
% --------------------------------------

Building on the analysis of swap-path choices, we next examine whether taking longer routes actually leads to better outcomes or simply adds costs. This question is central to MEV strategy design. Using transaction-level data, we compare path length with profit outcomes and capital efficiency across builders and tokens.

The correlations in Fig.\ref{fig:pearson-suite}\,(a) are all very close to zero for both \bd{48Club} and \bd{Blockrazor}, indicating that, on average, adding more swaps does not systematically increase or decrease profit per swap. The scatter plot in panel (d) reinforces this: \bd{48Club}’s profitable routes are concentrated in two- and three-swap bundles, with profit per swap quickly tapering off as paths extend; \bd{Blockrazor}’s points hug the horizontal axis across all path lengths, showing little incremental gain from complexity. At the same time, efficiency distributions in panel (b) diverge sharply. \bd{48Club}’s log\textsubscript{10} efficiency scores cluster roughly two to three orders of magnitude to the right of \bd{Blockrazor}’s, meaning that the same unit of deployed capital generates far more net profit for \bd{48Club}. The correlation matrix in panel (c) shows that core efficiency metrics (swap\_count, profit\_per\_swap, efficiency\_score, profit\_to\_fee\_ratio) are almost perfectly aligned internally, but only weakly connected to token-specific profit rates, suggesting that builder edge stems from execution quality rather than any single asset lane.  

These findings indicate that path length alone is not a driver of success. Strategic route selection, superior pricing models, and timely information matter far more than sheer complexity. For protocol design, this implies that modest hop limits or incentives against overly long routes could curb wasted gas and reduce externalities without harming overall MEV efficiency, while still allowing sophisticated builders to capture genuinely profitable opportunities.

\begin{figure}[!ht]
    \centering
    \subfigure[Transactions]{%
    \includegraphics[width=0.45\linewidth]{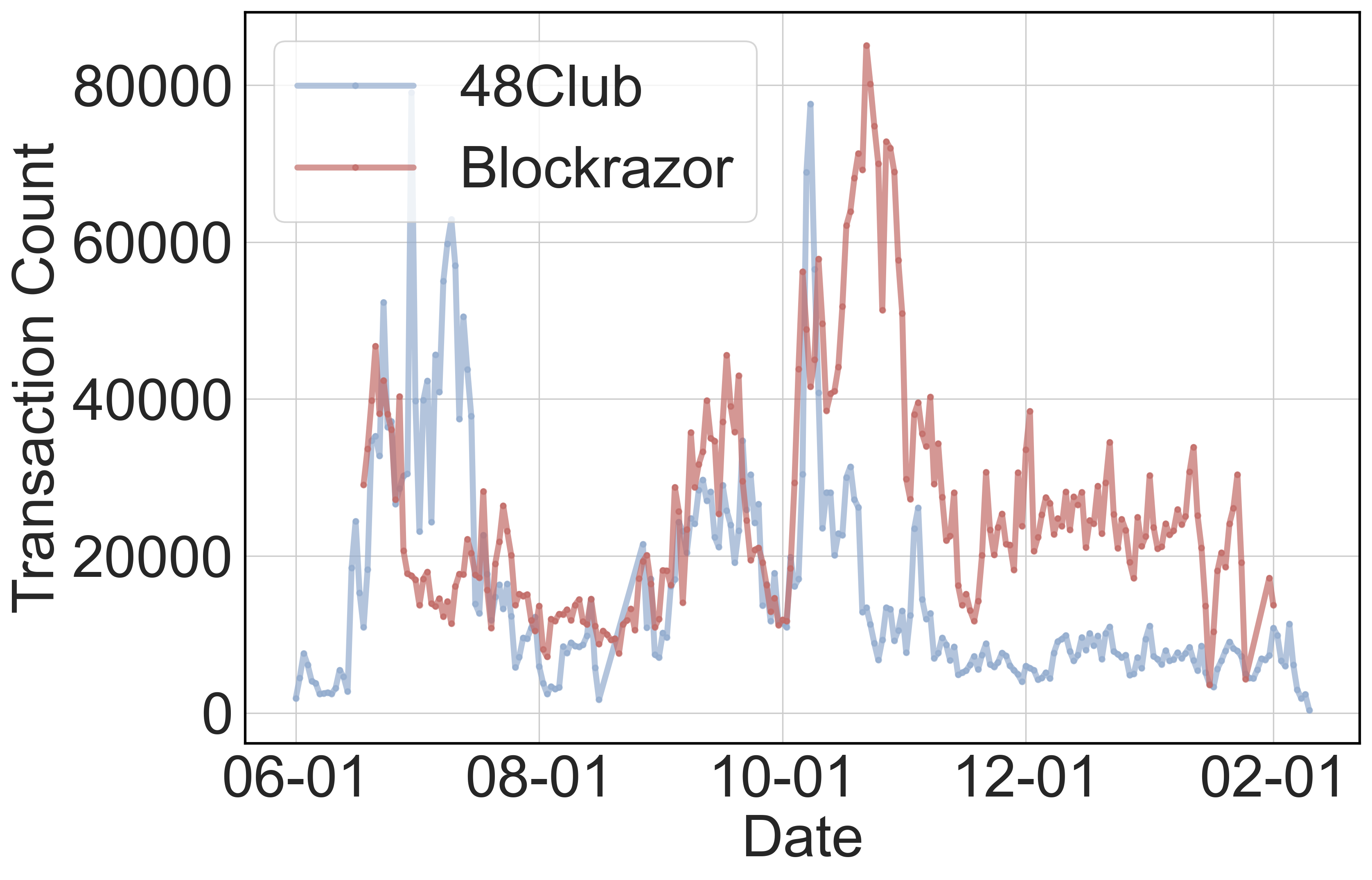}
    \label{fig:mk-tx}
    }
    \hfill
    \subfigure[Total profit]{%
    \includegraphics[width=0.45\linewidth]{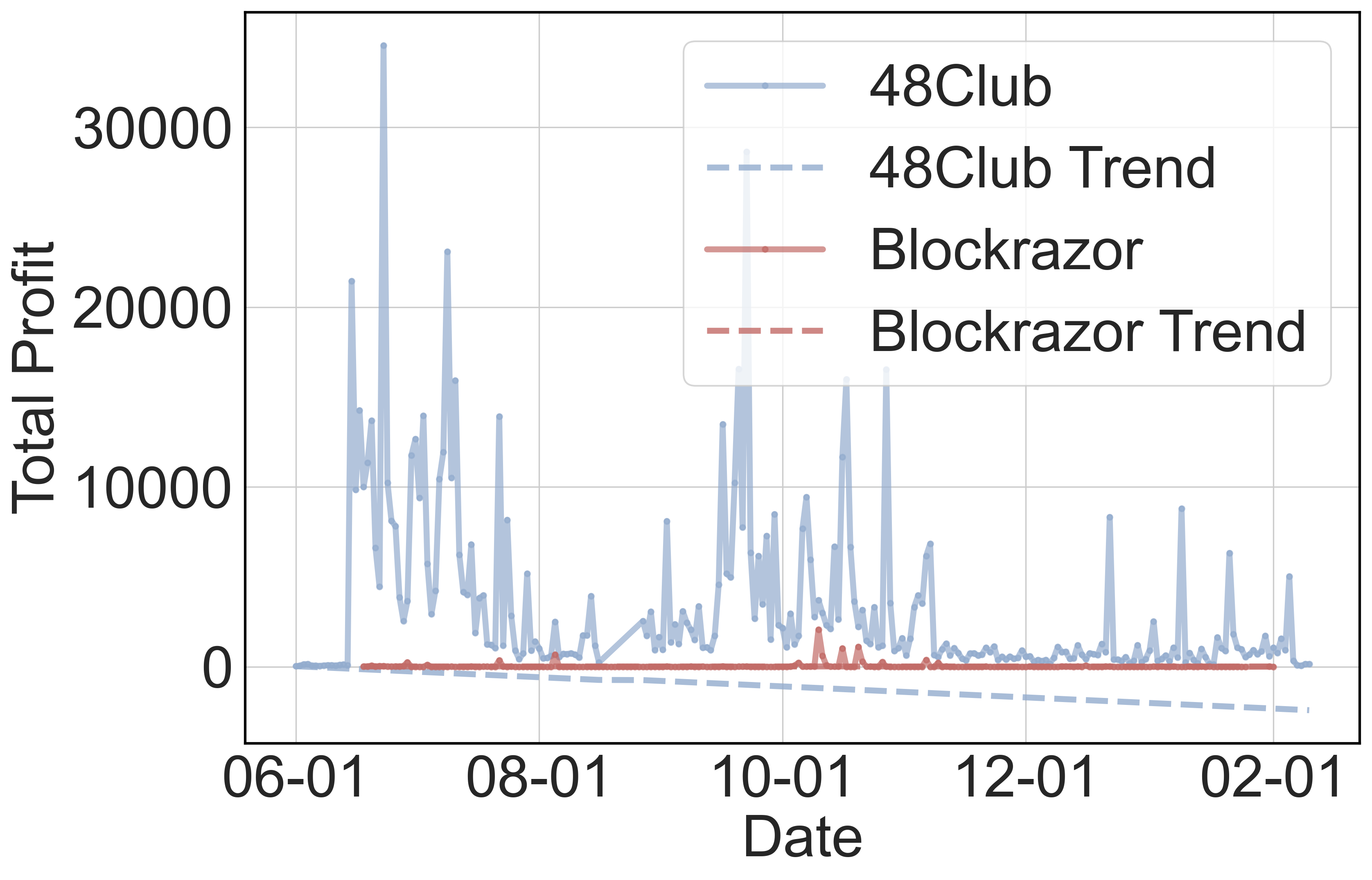}
    \label{fig:mk-total}
    }
    \subfigure[\texttt{WBNB}]{%
    \includegraphics[width=0.45\linewidth]{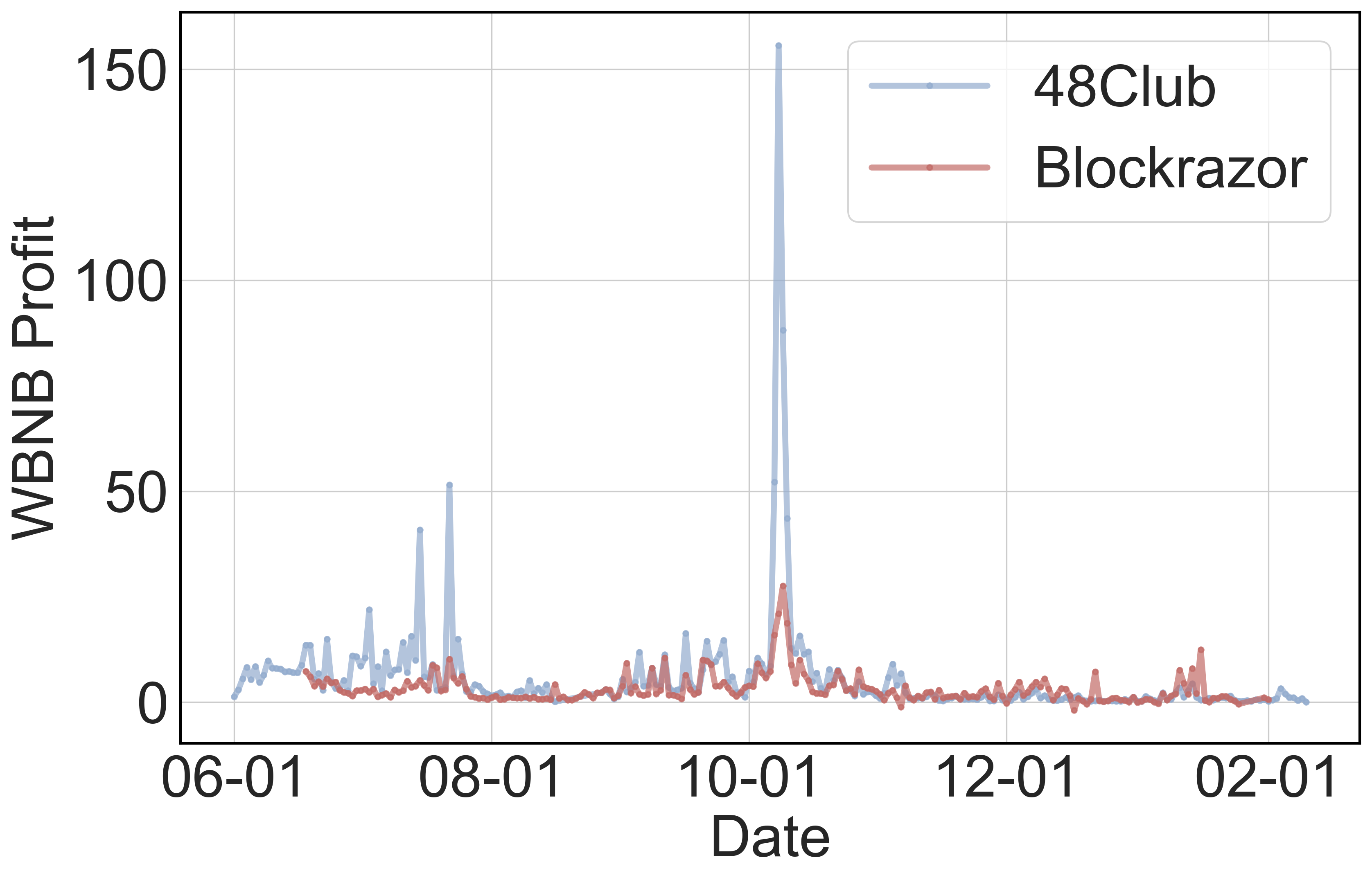}
    \label{fig:mk-wbnb}
    }
    \hfill
    \subfigure[\texttt{USDT}]{%
    \includegraphics[width=0.45\linewidth]{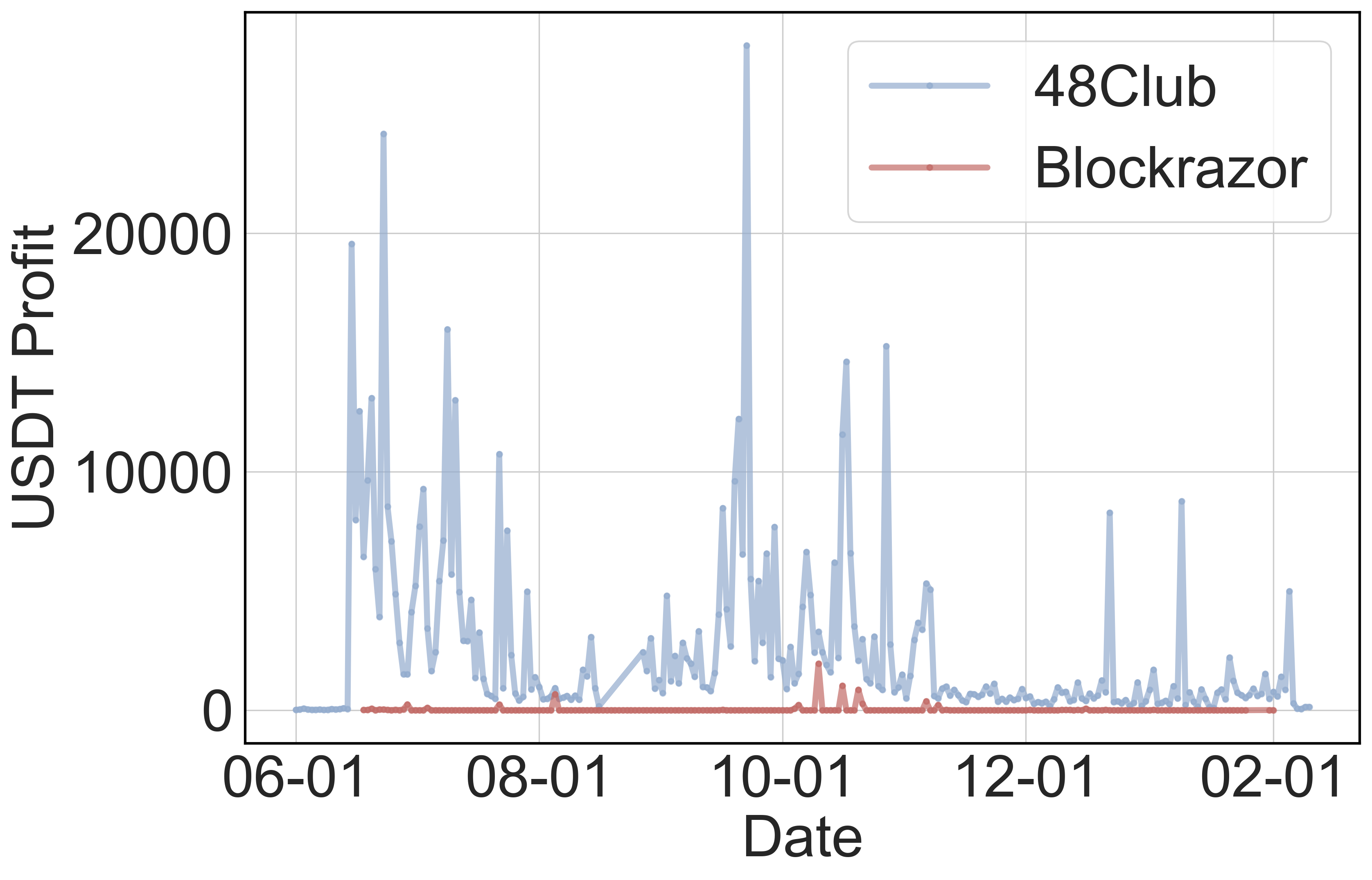}
    \label{fig:mk-usdt}
    }
    \hfill
    \subfigure[\texttt{USD1}]{%
    \includegraphics[width=0.45\linewidth]{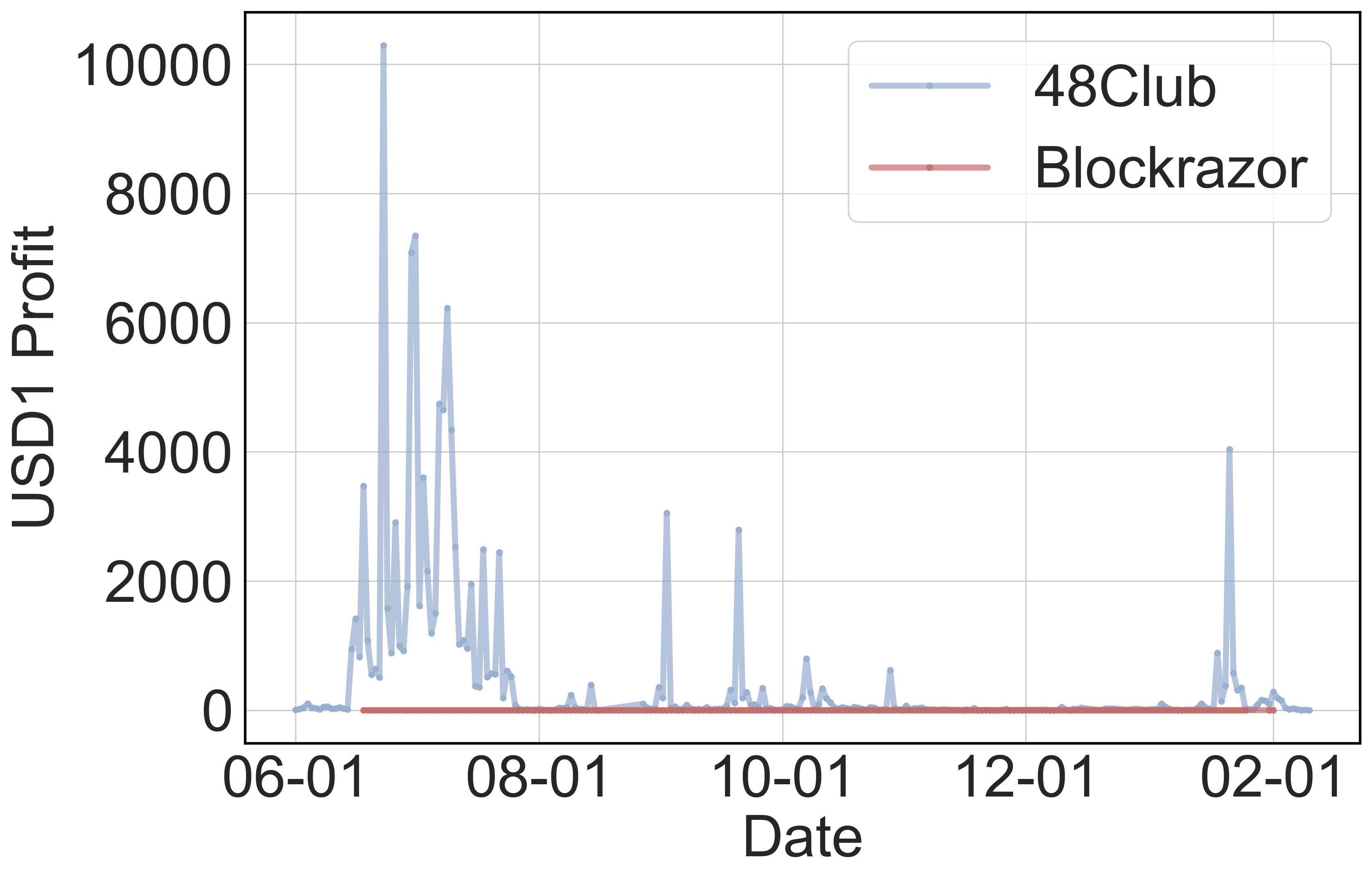}
    \label{fig:mk-usd1}
    }
    \hfill
    \subfigure[\texttt{USDC}]{%
    \includegraphics[width=0.45\linewidth]{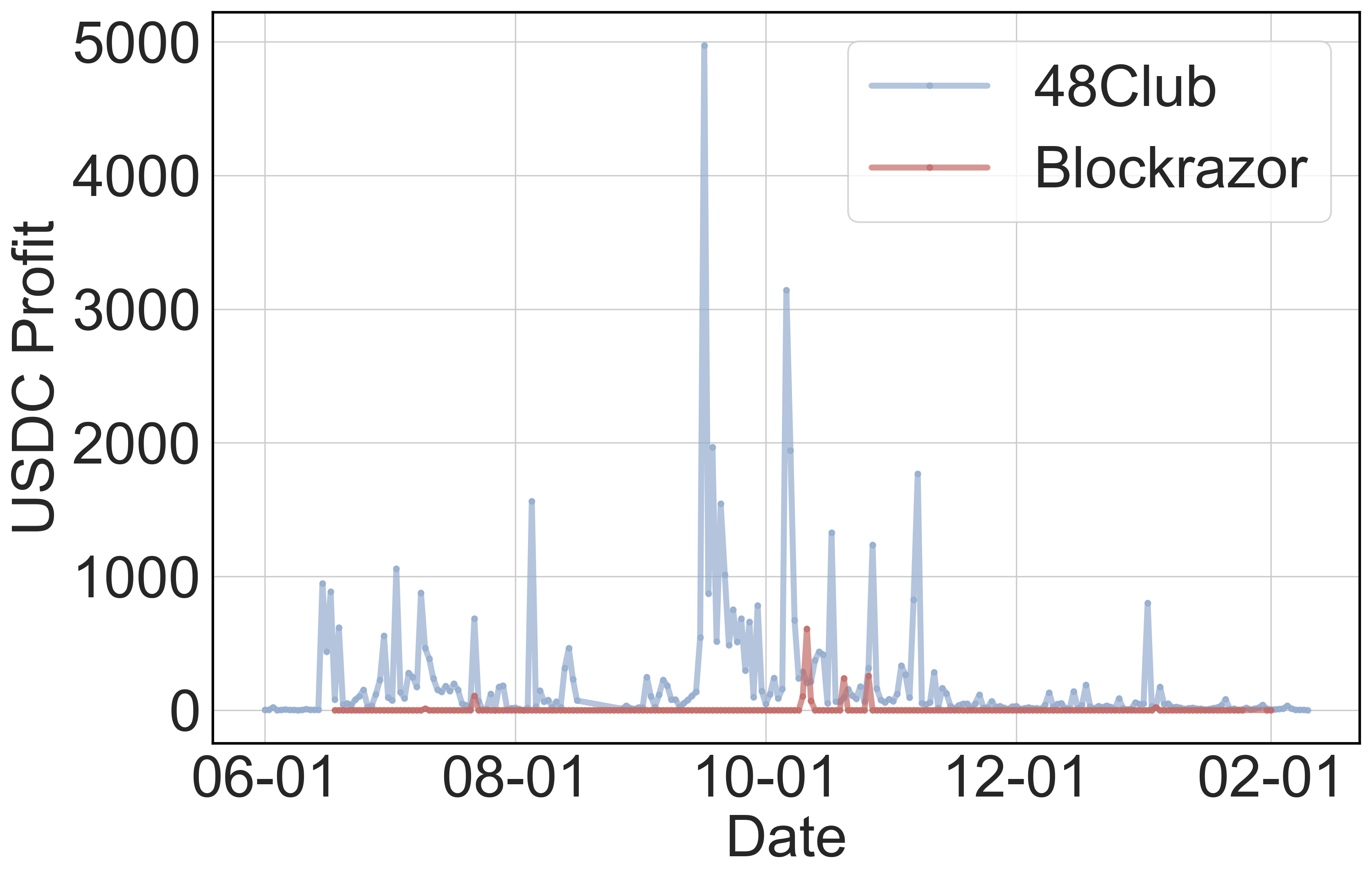}
    \label{fig:mk-usdc}
    }
\caption{Trends in MEV activity and profits on BSC. Blockrazor later catches up in volume, but 48Club leads in profit.}
    \label{fig:mk-trends}
\end{figure}

% --------------------------
\subsection{Activities/Profits Over Time (Fig.\ref{fig:mk-trends})}
\label{subsec:mann-kendall}
% --------------------------

To examine how builder activity and profits evolve over time, we track daily metrics across our April~1,~2025 to February~28,~2026 observation window and apply a non-parametric trend test that does not assume normality. Builder-operated arbitrage enters the sample from June onward, so the visible trend lines begin only after that point.

The contrast between volume and profit is striking. In Fig.\ref{fig:mk-tx}, \bd{48Club} drives the early wave of activity, ramping from a few thousand transactions per day in early June to peaks around 60{,}000 by mid-summer. From September onward, \bd{Blockrazor} catches up and at times exceeds 80{,}000 transactions per day, briefly overtaking \bd{48Club} in raw volume. Yet Fig.\ref{fig:mk-total} shows that \bd{48Club} continues to generate the dominant share of profit, with repeated spikes up to tens of thousands of WBNB-equivalent units. \bd{Blockrazor} remains close to the horizontal axis and its trend line stays zero.

Token-level panels reveal how this profit gap emerges. \texttt{WBNB} profits (Fig.\ref{fig:mk-wbnb}) show short-lived bursts for both builders but no persistent upward trend. \texttt{USDT} (Fig.\ref{fig:mk-usdt}) produces frequent and sizeable profit spikes for \bd{48Club} throughout the period, while \bd{Blockrazor} captures only thin traces. \texttt{USD1} (Fig.\ref{fig:mk-usd1}) shows clustered profit episodes that align with protocol, followed by long quiet stretches. \texttt{USDC} (Fig.\ref{fig:mk-usdc}) exhibits occasional spikes but overall remains the least profitable lane, with low median returns for builders.

These patterns indicate that MEV on BSC is shaped by both builder strategy and token microstructure. \bd{Blockrazor} eventually matches or exceeds \bd{48Club} in transaction count, but \bd{48Club} converts order flow into profit far more effectively, particularly in \texttt{USDT} and \texttt{USD1} pools. Profitability is therefore not just volume-driven but strongly token-dependent.

\begin{figure}[ht]
  \centering
  \subfigure[Centralization risk scores]{
    \begin{minipage}{0.46\linewidth}
      \centering
      \includegraphics[width=\linewidth]{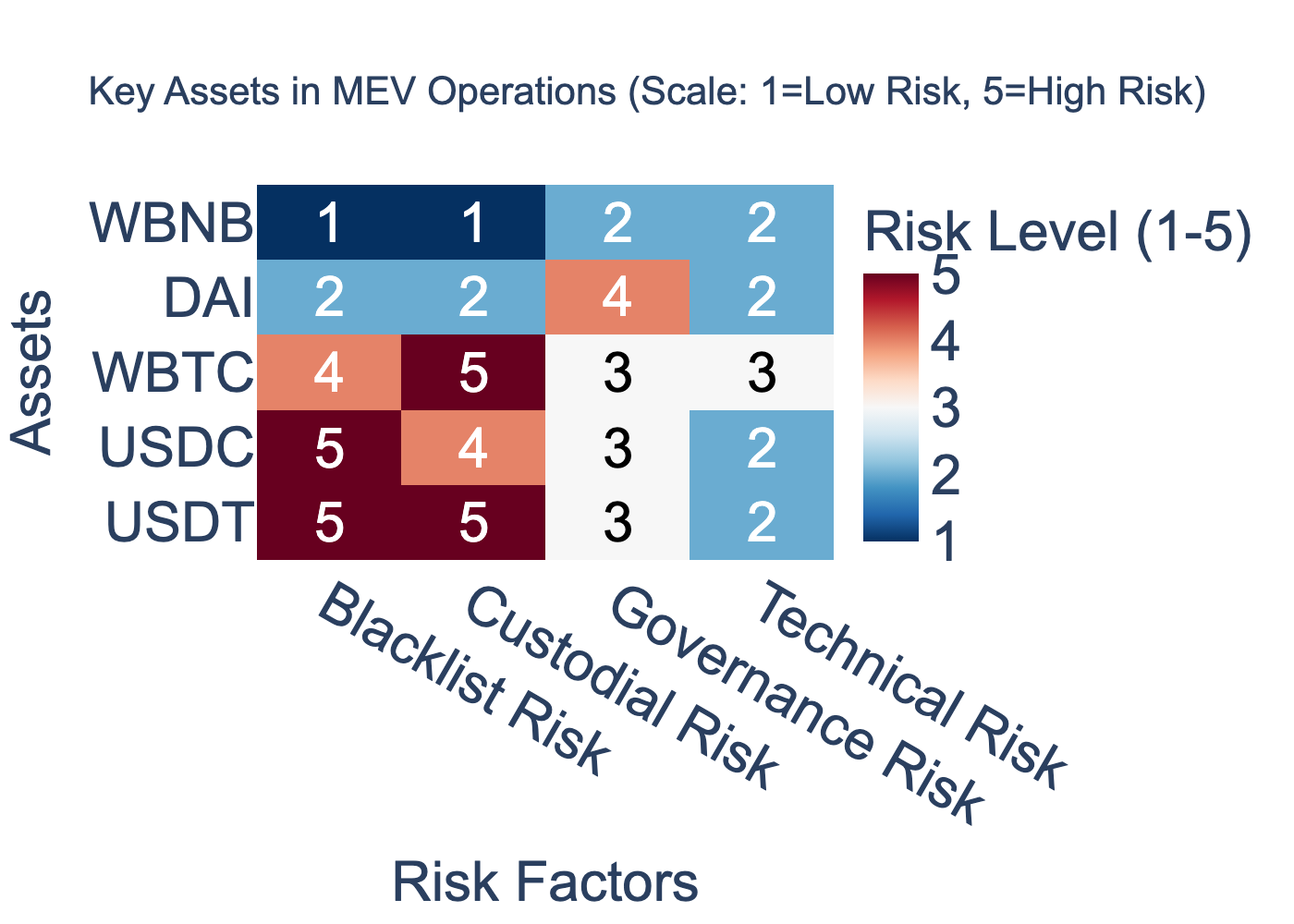}
      \label{fig:centralisation-risk-matrix}
    \end{minipage}
  }
  \subfigure[Share of routing volume]{
    \begin{minipage}{0.46\linewidth}
      \centering
      \includegraphics[width=\linewidth]{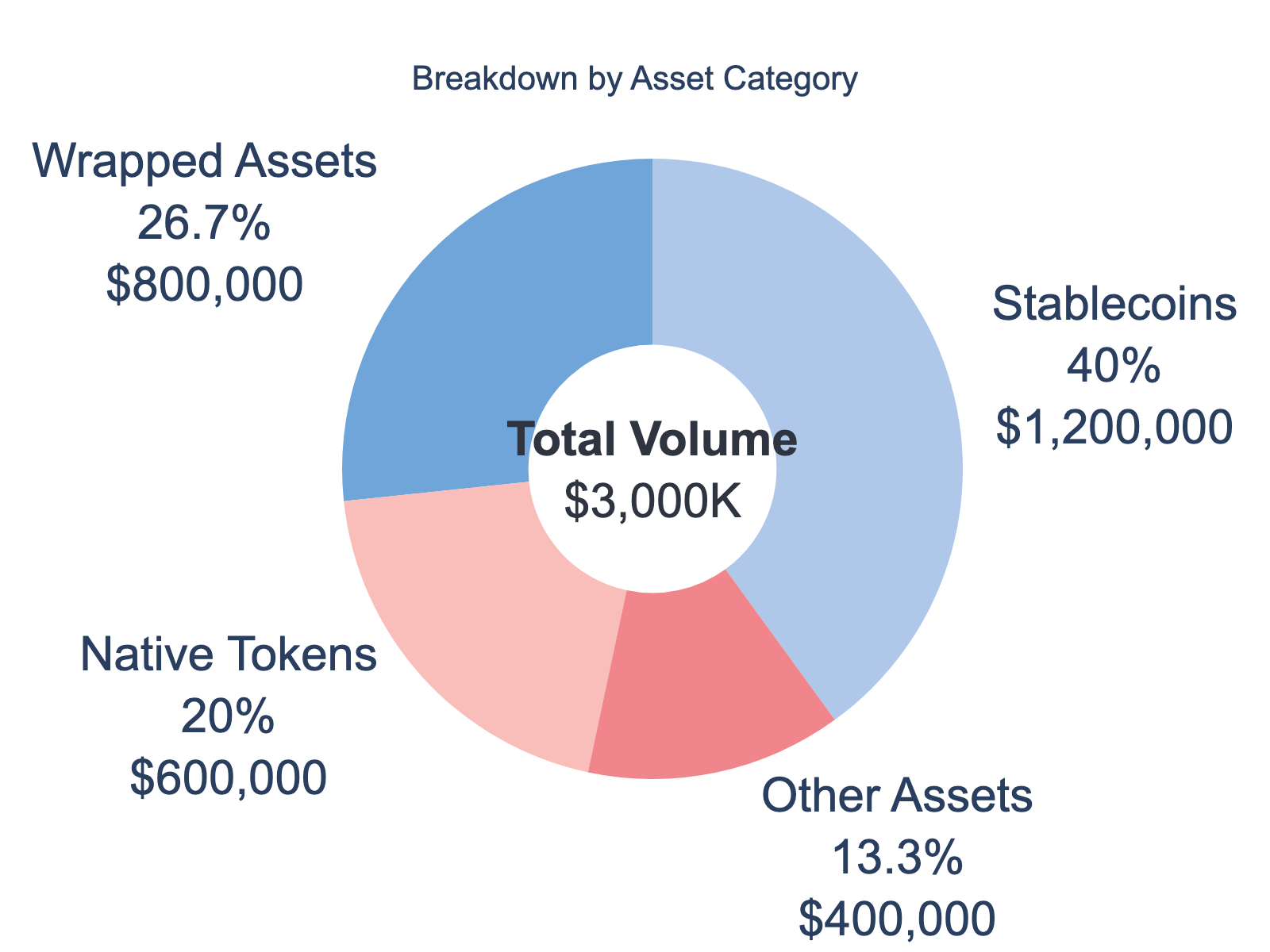}
      \label{fig:mev-asset-volume}
    \end{minipage}
  }
  \caption{Centralization in MEV assets. Stablecoins and wrapped assets are the most exposed.}
  \label{fig:mev-asset-risk}
\end{figure}

% --------------------------
\subsection{Centralization in MEV Assets (Fig.\ref{fig:mev-asset-risk})}
\label{subsec:centralisation-risk}
% --------------------------

We also measure the degree of centralization in MEV assets by combining two perspectives: the intervention risks each token carries and the share of routing volume it supports.

We assign each asset a normalized ‘centralization risk’ score in [0,1] based on three binary features: (i) whether the issuer or bridge contract can unilaterally freeze balances, (ii) whether redemptions depend on a single custodian or multisig, and (iii) whether the asset inherits risk from an external chain (e.g., via wrapped BTC). We then take the average across these features to obtain the score. 

The results (Fig.\ref{fig:centralisation-risk-matrix}) show that \texttt{USDT} and \texttt{USDC} score the highest on blacklistability and custodial dependence, while WBTC inherits similar fragility through bridge custodians. \texttt{DAI} is safer but still relies on governance and real-world collateral. By contrast, \texttt{WBNB} is the least risky as a natively wrapped, non-blacklistable token.  

Despite this spectrum of risk, most MEV activities continue to flow through the riskier assets. Stablecoins and wrapped tokens account for about two-thirds of routing volume, meaning that builders systematically rely on intermediates that can be censored or frozen. Even when final settlement occurs in \texttt{WBNB}, the majority of liquidity paths remain dependent on centralized rails, leaving the system fragile to external intervention (Fig.\ref{fig:mev-asset-volume}).

% --------------------------
\subsection{Asset Flows in MEV Operations (Fig.\ref{fig:mev-sankey})}
\label{subsec:asset-flow}
% --------------------------

We also analyze how MEV transactions actually move by tracing their origins, intermediate assets, and settlement venues. Most flows start from MetaMask, CEX hot wallets, or aggregators, then route through \texttt{USDT}, \texttt{USDC}, and \texttt{WBNB} before settling mainly on PancakeSwap and Uniswap V3.   

We trace token flows up to a bounded number of hops ($k = 4$) from each arbitrage transaction by following Transfer events that leave builder-linked contracts and DEX pools. Addresses are then coarsely bucketed into categories (CEX hot wallets, EOAs using MetaMask, aggregators, and other contracts) based on known labels and heuristics. Flows that cannot be classified within $k$ hops are grouped under ``Other/Unknown'' and remain a small fraction of total volume.

The results confirm earlier observations: high profits are not linked to long or complex paths, but to a narrow set of high-liquidity pairs and venues. Liquidity is funneled through a small number of assets, many of them censorable stablecoins, making MEV both highly profitable and structurally fragile.

\subsection{Answers from Data}

\begin{figure}[t]
  \centering
  \includegraphics[width=0.99\linewidth]{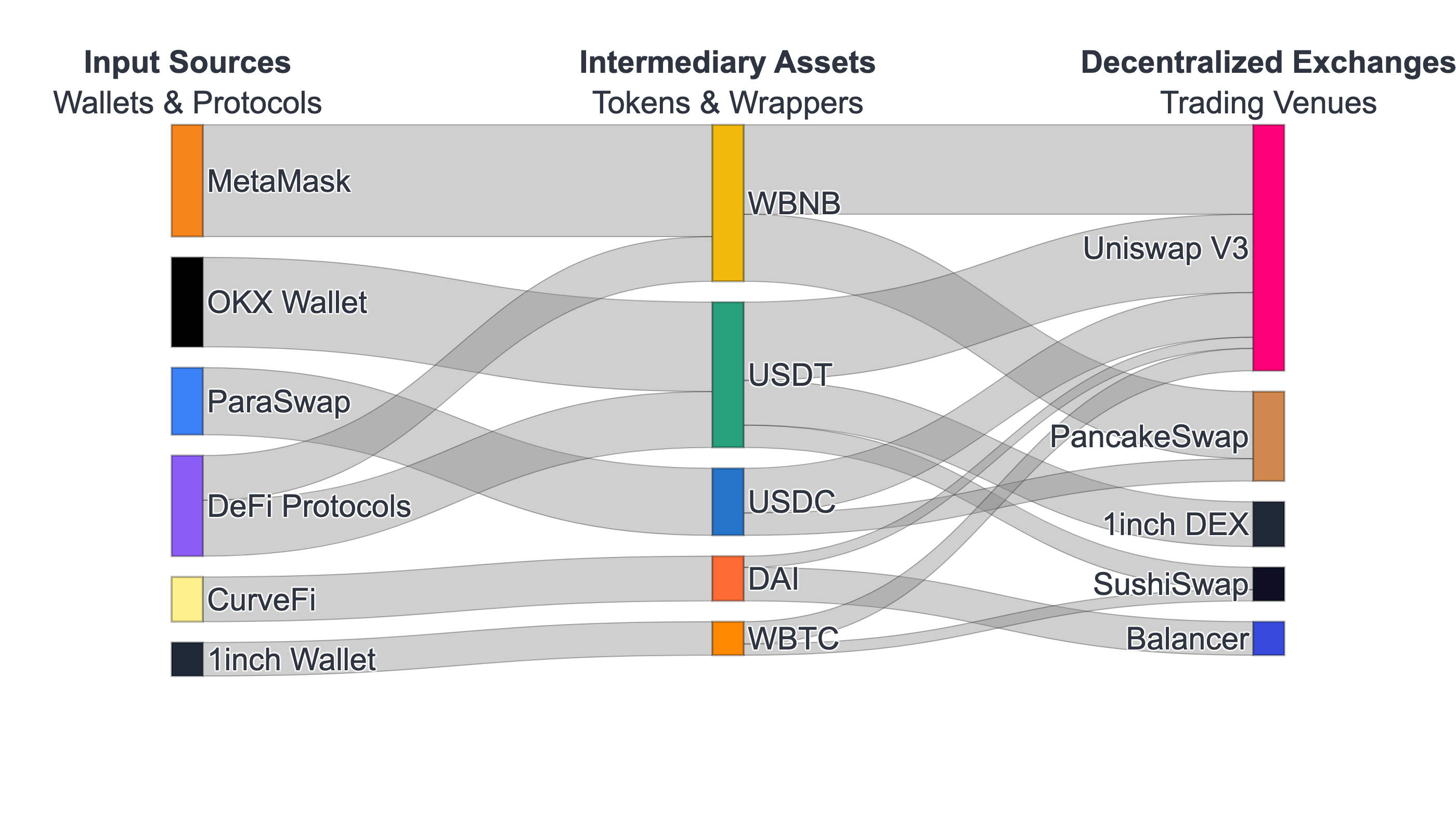}
  \caption{Asset flows. Most routes pass through \texttt{USDT}, \texttt{USDC}, and \texttt{WBNB} and settle on PancakeSwap or Uniswap V3.}
  \label{fig:mev-sankey}
\end{figure}

Across the preceding measurements, a single pattern recurs: profits are concentrated in a narrow set of builders, tokens, routes, and venues. The dominant gains accrue to 48Club, especially through stablecoin-heavy opportunities centered on \texttt{WBNB}, \texttt{USDT}, and \texttt{USD1} (Figs.~\ref{fig:chisquare_combined} and \ref{fig:mk-trends}). Most successful arbitrage uses short two- or three-swap paths; longer paths add complexity but little incremental return (Figs.~\ref{fig:ks_combined} and \ref{fig:pearson-suite}). The asset-flow analysis further shows that these routes repeatedly pass through a small liquidity backbone dominated by PancakeSwap, Uniswap V3, and a few high-liquidity token rails (Fig.~\ref{fig:mev-sankey}).

The data imply that BSC MEV is not merely concentrated by actor. It is also concentrated by timing, venue, and asset microstructure. Builders that control the fastest order flow and the deepest stablecoin routes repeatedly harvest the same classes of opportunities, while smaller participants are largely confined to residual, low-margin events.

%=================================================
\section{STRUCTURAL RISKS IN BSC PBS}
\label{sec-central}
%=================================================

We now discuss the centralization (\S\ref{subsec-centralized}), censorship (\S\ref{subsec-censorship}), MEV inequality (\S\ref{subsec-inequality}) and latency impact (\S\ref{subsec-latency}) in BSC.

\subsection{More Centralized?}
\label{subsec-centralized}

Ethereum PBS was designed to keep the builder market open, yet in practice only a few large builders dominate \cite{wang2025private,wu2025competition,mancino2025decentralization,burian2024mev,bahrani2024centralization,gupta2023centralizing}. BSC’s design amplifies this trend even further. A short block time and a small validator set make the network more centralized. We now examine how this happens.

\vspace{3pt}
\noindent\textbf{Shorter block times affect builder behavior.}  
With little time for multi-round negotiation, builders keep a persistent, low-latency link to the scheduled proposer and submit compact bids early, often bundling an immediate payload hand-off (i.e., a single round of bid+full block delivery). Because the proposer must execute the transactions to derive a valid header before signing, builders cannot hide contents behind a header. After bid acceptance, they must transmit the full transaction list quickly, which further compresses their latency budget.

This timing pressure pushes builders toward pre-positioned, high-frequency strategies: short and low-hop cycles (e.g., triangular or stablecoin routes), deterministic routing, and automated pipelines that minimize IO and validation overhead. Order flow is sourced through direct channels rather than relying on slow propagation.  

\vspace{3pt}
\noindent\textbf{Independent searchers face a structural disadvantage.}  Our on-chain measurements only capture the part of MEV that is ultimately retained by builders and redistributed to proposers. In practice, many opportunities are first discovered by off-chain searchers who submit bundles via private auctions, and builders keep only the share they negotiate in these auctions. From the chain’s perspective, however, all such cycles are executed by builder-owned contracts and appear as “builder-driven” MEV. Our data show that any MEV that reaches the chain must pass through a very small number of whitelisted builders, giving them substantial control over which searchers are served and which bundles are forwarded to proposers.

The 3-second block interval and the whitelisted builder interface still put independent searchers at a disadvantage: they must negotiate with dominant builders under tight latency and information asymmetry, instead of interacting directly with proposers. First, they lack the time to detect and react to price gaps before builders act. By the time a searcher broadcasts to the mempool, a builder with direct access has already packaged the opportunity. Second, validators prefer reliable, high-paying builders, not random searchers with late or uncertain bundles. This contrasts with Ethereum’s more open market, where searcher participation remains observable when competition per opportunity is low, although validators capture more surplus as competition intensifies \cite{mamageishvili2024searcher}. Thus, most searcher transactions on BSC either lose the race or arrive only after the builder has captured the profit.

\vspace{3pt}
\noindent\textbf{Validator–builder ties.}  
On top of the whitelisted builder set, PoSA is highly concentrated: at any time only about 40–45 validators are active, and a handful of large staking pools routinely sign the majority of blocks. This makes it easier for top builders and validators to form long-lived bilateral relationships than in Ethereum’s much larger validator set.

Validators therefore form stable, long-term ties with top builders who consistently deliver profitable blocks. This relationship further reduces market openness: once a validator–builder channel is established, it is difficult for new entrants to compete. Compared to Ethereum’s larger, more diverse validator pool and public relays, BSC creates a narrower and more exclusive MEV market.  

Our proposer-share analysis (\S\ref{subsec:proValid}) confirms this dynamic: \bd{48Club} distributes sizable proposer payouts across several tokens, helping sustain broad validator alignment, whereas \bd{Blockrazor} concentrates payouts in a narrower swapped-\texttt{USDT} lane. These strategies show that validator incentives are not neutral, but help determine which builders dominate and which fade away.

\vspace{3pt}
\noindent\textbf{High operational bar.}  
Even worse, because of both whitelist admission and strict timing, only builders with high-performance infrastructure can survive. The need for ultra-low latency networking, optimized mempool access, and pre-arranged transaction flows raises the entry bar much higher than on Ethereum. Over time, this may reinforce concentration, with only a handful of builders able to stay competitive.

\vspace{3pt}
\noindent\textbf{Narrow MEV strategy space.}  
Short block times also shape which MEV strategies remain viable. Simple triangular swaps or two-hop stablecoin routes dominate, since they can be executed quickly and reliably. More complex strategies are less practical under the three-second window, such as long multi-hop arbitrage, slow liquidations, or oracle-based attacks. This filtering effect narrows the diversity of MEV. Only fast, low-hop opportunities are favored by builders.  

\vspace{3pt}
\noindent\textbf{Implications for rollups/L2s.}  
BSC’s short block interval not only centralizes MEV on L1 but also pushes pressure upward to L2 protocols (e.g.,  \cite{solmaz2025optimistic,ferreira2024rolling}). Rollups and sidechains that settle on BSC must adapt to very fast confirmation windows. On one hand, this shortens the time to finality for bridging and settlement. On the other, it reduces the reaction time for sequencers and bridge operators to guard against MEV, such as sandwiching or oracle manipulation~\cite{zhou2023sok,zhou2021high,eskandari2021sok}. Faster settlement thus improves user experience but increases risk, as defenses must react within just a few seconds.  

\subsection{More Censorship?}
\label{subsec-censorship}

Censorship refers to selectively including or excluding transactions from blocks. On Ethereum, it rose to prominence after the Tornado Cash sanctions, where relays and validators began following OFAC listings~\cite{wang2023first,xiong2024global}. Because PBS centralizes inclusion power, the mechanism amplifies such risks~\cite{wadhwa2025aucil}. On BSC, these risks manifest more sharply. We examine four concrete dimensions.

\vspace{3pt}
\noindent\textbf{Builders filter transactions.}  
On BSC, proposers earn a share of builder profits (Fig.~\ref{fig:validator_profit_combined}), giving them direct financial motivation to prioritize bids from whitelisted builders instead of public mempool traffic. Since only approved builders may submit bids, the block body is already pre-filtered long before it reaches the proposer.  Our traces show that builders routinely drop low-fee or non-arbitrage transactions from their bundles when these reduce net profit, even if those transactions are otherwise valid.   Validators reinforce this by enforcing a minimum 3~Gwei floor and occasionally inserting their own 1~Gwei transactions through private \emph{inclusion lists}, bypassing normal mempool competition. The combined effect is a narrow and economically curated transaction set, where exclusion is not exceptional but a routine consequence of bundle construction.

\vspace{3pt}
\noindent\textbf{Network favors insiders.}  
Most validators operate behind private sentry nodes and maintain persistent peering channels with major builders. These low-latency, authenticated channels receive privileged ordering priority over the public P2P network.  
Our measurements show that public mempool transactions often arrive dozens of milliseconds later than builder-submitted bundles, enough to render them uncompetitive in a 3-second block competition.  Privacy-enhancing transactions suffer the most: even without explicit filtering, they are rarely propagated through these private overlays. This form of \textit{soft} censorship emerges not from protocol-level rules but from networking topology.

\vspace{3pt}
\noindent\textbf{Stablecoins add censorship.}  
Censorship is not limited to block inclusion. The dominant assets on BSC—\texttt{USDT} and \texttt{USDC}—retain issuer-level freeze and blacklist functions. These mechanisms are actively used across chains (e.g., post-Multichain exploit freezes), and BSC inherits these constraints.  
A transfer included by a validator may still revert at the contract level if the address is frozen. Builders, whose rewards depend on reliable execution and payout, have an incentive to avoid routing arbitrage through addresses or pools associated with freeze risk. This leads to \emph{asset-level} censorship: certain flows are never attempted because the token contract’s rules introduce asymmetric execution risk.

\vspace{3pt}
\noindent\textbf{Private relays as gatekeepers.}  
Although BEP-322 does not mandate relays, practice reveals a de facto relay ecosystem.   Many BSC validators connect to commercial RPC or relay infrastructures such as \bd{Bloxroute} to reduce propagation delays.  
These private channels apply their own onboarding, traffic shaping, and filtering policies~\cite{lyu2025demystifying,zhang2024breaking}.  
In several cases, they have already demonstrated selective suppression of sanctioned or “flagged’’ flows on Ethereum, and nothing prevents the same templates from being applied to BSC.   For ordinary users relying on the public mempool, this produces a two-tier network: fast-path private channels for builders and approved clients, and a slow, congestion-prone path for everyone else.  Filtering thus becomes a structural property of relay access rather than a discretionary validator choice.

\subsection{MEV Inequality in BSC}
\label{subsec-inequality}

\noindent\textbf{Block-building inequality.}
Ethereum-based analyses~\cite{gupta2023centralizing,yang2025decentralization,wang2025private} quantify inequality using metrics such as true-value dispersion, second-best builder estimates, and inequality loss. Although our measurements do not replicate the same formal metrics, the empirical patterns on BSC indicate greater inequality in practice. The profit paths we observe---dominated by two- and three-hop stablecoin arbitrage---produce a narrow and deterministic opportunity structure. Builders able to secure the fastest private flow repeatedly capture nearly identical opportunities across consecutive slots, while slower builders rarely observe meaningful profit.

This dynamic collapses the effective competition set. In many epochs, proposers receive only one viable bid, or face a gap of more than an order of magnitude between the top builder and the next-best competitor. Thus, BSC proposers do not operate in a genuinely competitive builder market. Realized MEV is shaped less by auction pressure than by the pricing strategy of dominant builders. Validator rotation does not remove this imbalance: the proposer changes, but the same narrow builder set continues to control the profitable order flow and determine the credible bid range. The result is a structurally high level of block-building inequality.

\vspace{3pt}
\noindent\textbf{Latency as a source.}
Although our measurements do not directly capture network-level latency, several on-chain behaviors indirectly reveal how timing shapes MEV outcomes on BSC. Most notably, arbitrageable price deviations on \texttt{USDT}, \texttt{USD1}, and \texttt{WBNB} pools persist for only 150--400\,ms, and builder-submitted transactions consistently occupy the earliest intra-block positions, while searcher-originated transactions almost never preempt a builder. These patterns indicate that MEV extraction on BSC operates on a sub-second timescale, even though block intervals are 3 seconds.

This divide between block time and opportunity time is central.  
A builder that receives order flow 30\,ms earlier than its competitors can compute and submit a route while the deviation is still profitable; a builder that arrives 80--120\,ms later typically observes a market that has already been arbitraged.  Expected returns decay sharply after the first 100\,ms and fall below gas cost around 200\,ms (Fig.\ref{fig:slot-inequality}), creating a near-binary success profile: early arrivals capture the entire opportunity, while later arrivals observe stale quotes.

This latency sensitivity is amplified by BSC’s whitelisted PBS design.  
Once the proposer accepts the first valid bid, slower builders cannot inject another bundle, even if their route is more profitable. Execution follows a winner-takes-all dynamic.  
By contrast, Ethereum’s 12-second slots allow late-arriving transactions to be included  without competing in a sealed bid race, softening the effect of small latency skew and giving searchers a wider participation range.

We conclude that MEV extraction on BSC operates at the millisecond scale.  Block time determines when results are finalized; opportunity time determines who captures the profit.  The gap between these two timescales introduces structural inequality: actors with privileged and low-latency connectivity will win, while others (i.e., searchers or smaller builders) rarely reach the proposer within the profitable window.

\begin{figure}[t]
\centering
\resizebox{\linewidth}{!}{%
\begin{tikzpicture}[>=stealth]

  % --- Time axis ---
  \draw[thick,->] (0,0) -- (7,0) node[below right]{\small Time scale};

  % ===== Sub-second latency (Δ_lat) =====
  \draw[thick] (1,0.1) -- (1,-0.1);
  \node[below] at (1,-0.1) {%
    \begin{tabular}{c}
      \scriptsize Sub-second latency \\
      \scriptsize $\Delta_{\text{lat}}\approx 10^2\text{ ms}$
    \end{tabular}
  };

  % ===== 3s slot (H_BSC) =====
  \draw[thick] (3.3,0.1) -- (3.3,-0.1);
  \node[below] at (3.3,-0.1) {%
    \begin{tabular}{c}
      \scriptsize 3 s slot \\
      \scriptsize $H_{\text{BSC}}\approx 3\text{ s}$
    \end{tabular}
  };

  % ===== 12s slot (H_ETH) =====
  \draw[thick] (6,0.1) -- (6,-0.1);
  \node[below] at (6,-0.1) {%
    \begin{tabular}{c}
      \scriptsize 12 s slot \\
      \scriptsize $H_{\text{ETH}}\approx 12\text{ s}$
    \end{tabular}
  };

  % ===== Latency race window =====
  \draw[decorate,decoration={brace,amplitude=4pt}] 
    (0.3,0.6) -- (1.7,0.6)
    node[midway,above=4pt] {\scriptsize Latency race $[0,\Delta_{\text{lat}}]$};

  % ===== Ethereum extra coordination window =====
  \draw[decorate,decoration={brace,amplitude=4pt}]
    (1.7,1.2) -- (6,1.2)
    node[midway,above=4pt] {\scriptsize Extra coordination $(\Delta_{\text{lat}},H_{\text{ETH}}]$ on Ethereum};

  % ===== Red lost-window segment (H_ETH - H_BSC) =====
  \draw[very thick,red!70] (3.3,0) -- (6,0);

  % <<<<< --- Move label ABOVE the red line --- >>>>>
  \node[above=6pt, red!70!black] at (4.65,0) {\scriptsize Missing horizon $H_{\text{ETH}}-H_{\text{BSC}}$};

\end{tikzpicture}%
}

\caption{Time scales for MEV competition. Ethereum has a longer coordination window; BSC largely collapses it.}
\label{fig:slot-inequality}
\end{figure}

\subsection{Missing Horizons (Fig.\ref{fig:slot-inequality})}
\label{subsec-latency}

We also identify a structural constraint of BSC’s PBS that we call the \emph{missing horizon}---the portion of the block-building interval in which MEV competition \emph{could} occur but structurally cannot. This horizon exists on Ethereum but is effectively absent on BSC.

\vspace{3pt}
\noindent\textbf{The block-building horizon.}
For any MEV opportunity, block construction has two temporal regions. The first is a short \emph{latency race window}, roughly $\Delta_{\text{lat}}\approx 10^2$ ms, in which builders compete to deliver a profitable bundle first. Both Ethereum and BSC share this region.

The second region, specific to longer-slot designs, is the \emph{coordination window} $(\Delta_{\text{lat}}, H]$, where $H$ denotes the block-building horizon. In Ethereum’s 12-second slot, this window is nearly an order of magnitude longer than the latency interval. It allows builders to rebid, resubmit improved blocks, substitute transactions, or route flow to stronger builders through relays. This coordination phase helps equalize competition and limits latency-based monopolies.

\vspace{3pt}
\noindent\textbf{Missing horizon.}
On BSC, the block-building horizon is only $H_{\text{BSC}}\approx 3$\,s. After subtracting the sub-second latency window, almost no usable time remains. Proposers do not wait for secondary bids; builders cannot meaningfully revise or
replace blocks; and no coordination mechanisms (such as relays, sealed headers, or late substitution) can operate. Thus the effective coordination window on BSC is vacuous: $(\Delta_{\text{lat}}, H_{\text{BSC}}]\approx \varnothing$.

We define the gap $H_{\text{ETH}} - H_{\text{BSC}}$ as the \textit{missing horizon}--the region of MEV competition that exists on Ethereum but is structurally unattainable on BSC. This gap represents \emph{lost contestability}: opportunities that, on Ethereum, would have received multiple bids, undergone relay filtering, or been improved by slower builders, but on BSC are awarded solely according to arrival time.

%=================================================
\section{Further Discussion}
\label{sec-discu}
%=================================================

We extend our findings in this section.

\subsection{Why does BSC represent an extreme case of PBS centralization?}

Our findings show that BSC amplifies several structural tendencies inherent to PBS. Short slot times magnify the impact of latency advantages; whitelisting restricts builder entry and freezes market structure; private order flow channels concentrate opportunities; and the small PoSA validator set facilitates persistent proposer-builder ties. These properties interact \textbf{multiplicatively} rather than additively: each factor reinforces the others, producing a highly centralized MEV extraction pipeline.

Ethereum has similar pressures (e.g., private flow dominance, builder inequality, incentive distortions), but its larger validator set, relay architecture and permissionless builder market introduce countervailing forces that prevent extreme concentration. BSC lacks these balancing mechanisms.

\subsection{Mitigation for a Healthier PBS}

We outline a non-exhaustive set of strategies, focusing on changes that are compatible with BSC’s design.

\begin{packeditemize}
\item\textit{Introducing a lightweight relay layer.}
Unlike Ethereum, BSC relies on direct builder-proposer communication, which exacerbates latency advantages and facilitates exclusive relationships. A lightweight relay (not necessarily as complex as MEV-Boost) could provide sealed-header forwarding, defend against bid stealing, and decouple builders from proposers. Even a minimal commit-reveal step would dampen the advantage of privileged network positioning and allow proposers to evaluate multiple bids fairly within BSC’s 3-second slot.
    
\item\textit{Expanding the builder whitelist with transparent entry rules.}
The current whitelist restricts competition and entrenches existing builder dominance. Establishing transparent criteria for new entrants (e.g., liveness guarantees, performance, or hardware requirements) would lower the barrier for challenger builders. Periodic re-evaluation of whitelisted builders could further prevent long-term market capture and reduce the risk of proposer-builder collusion.

\item\textit{Opening partial order flow sources.}
Our measurements show that profitable arbitrage flow rarely reaches the public mempool. Introducing optional public order flow channels could diversify opportunity access and weaken the monopoly held by private RPC endpoints. Even a small fraction of publicly accessible flow would enable new builders to bootstrap, observe real competition patterns, and estimate expected true-value more accurately.

\item\textit{Enforcing proposer-side minimum-bid or diversity filters.}
Validators currently accept nearly all blocks from dominant builders, even when no competitive bids exist. Simple proposer-side policies (e.g., requiring bids from at least two builders when available, or imposing minimum-revenue thresholds relative to slot-level MEV estimates) could reduce proposer dependence on a single builder and discourage uncompetitive bids. These mechanisms would also increase proposer revenue, which is disproportionately low on BSC due to 0-gas-price blocks.
\end{packeditemize}

\subsection{Limitations and Biases}
Our measurements focus on the two dominant builders whose arbitrage contracts are publicly identifiable on-chain. While smaller builders may also conduct MEV activities, their block share is negligible and does not affect our concentration results. The observation window covers Apr.,~2025 to Feb.,~2026, while builder-operated arbitrage becomes visible only from late May and mid-June~2025 onward; earlier activity in the window remains largely searcher-driven and is therefore not directly comparable.

We measure only on-chain observable activity. Private RPC order flow between searchers and builders cannot be inspected directly, and we do not observe off-chain profit splits. However, these limitations do not affect block-level dominance, as builder identities remain visible regardless of transaction origin. Accordingly, our statements about searcher competitiveness should be interpreted as structural observations about access to block construction rather than precise estimates of searcher margins.

Our analysis focuses on arbitrage-based MEV, which dominates builder-executed activity during the observed period. Other MEV forms (e.g., backrunning or liquidations) occur far less frequently and do not materially influence concentration metrics.

Finally, although our labeling captures over 95\% of builder-linked activity, a small number of short-lived contracts may remain unlabeled. Such omissions would only reduce the measured activity of minor builders and therefore do not affect our conclusions on dominance, profit distribution, or market concentration.

%====================================================
\section{Extended Related Work}
\label{sec-rw}
%====================================================

Core studies were addressed earlier. We add an extension.

\vspace{3pt}
\noindent\textbf{MEV in different forms.}  
Beyond typical strategies of frontrunning, backrunning, or sandwiching~\cite{daian2020flash,zhou2021high,wang2022exploring,zhang2024front,zhang2025no}, MEV operates in a variety of other arbitrage types.
Non-atomic arbitrage across DEXes~\cite{heimbach2024non} accounted for more than a fourth of Ethereum’s largest DEX volume after the Merge, with just eleven searchers responsible for over 80\% of the identified \$132B flow.
CEX–DEX arbitrage~\cite{wu2025measuring} was measured at over \$230M, where price gaps between centralized and decentralized exchanges were exploited, again with only a handful of searchers capturing the majority.
Cross-chain arbitrage~\cite{oz2025cross} reported \$868M of volume within a year, dominated by a few large players.
Cross-rollup arbitrage~\cite{gogol2024cross} occurred 2–3$\times$ more often than on the Ethereum mainnet, though with lower trade sizes, and persisted across 10–20 blocks.

Time also creates new MEV opportunities. Ethereum’s move to PoS enabled \textit{waiting games} \cite{oz2023time}, where validators strategically delay block proposals to capture additional arbitrage. Recent analyses~\cite{fritsch2024mev} further show that auctioning off time advantages can boost validator profit without breaking consensus.  Schwarz et al.~\cite{schwarz2023time} model time-delay strategies in PBS explicitly and show that such timing games can persist even under competitive builder markets.

\vspace{3pt}
\noindent\textbf{MEV mitigation solutions.} We noted four types of countermeasures~\cite{heimbach2022sok,baum2022sok,yang2024sok}. 
Auction platforms (e.g., Flashbots, MEV-Boost, SUAVE) move transaction ordering off-chain to reduce fee wars but raise centralization and censorship concerns~\cite{wang2023blockchain,wahrstatter2024blockchain}. 
Time-based ordering enforces arrival-time fairness~\cite{gramoli2024aoab,yahyaoui2025mitigating,cachin2022quick,zhang2020byzantine,kelkar2023themis} at the protocol layer, though it risks latency races and stronger synchrony assumptions. 
Content-agnostic ordering hides transaction details via commit–reveal or hardware-assisted encryption~\cite{li2023transaction,babel2024prof,zarbafian2023aion}, limiting frontrunning but requiring extra trust or cryptographic support. 
Application-layer designs (e.g., batch auctions~\cite{zhang2025maximal}, redistribution~\cite{zhang2024rediswap}, or inclusion lists \cite{wadhwa2025aucil}) embed MEV-resistance directly in DeFi protocols, with inclusion lists constraining builders by enforcing minimum transaction sets to mitigate censorship.  

These countermeasures remain conceptually relevant for BSC’s PBS, but their effectiveness is constrained by PoSA’s design: auction platforms may further centralize and fairness-based ordering is harder to enforce under tight latency. Only application-level mitigations appear transferable.

\vspace{3pt}
\noindent\textbf{Decentralization of PBS.} Recent studies~\cite{yang2025decentralization,wang2025private,wu2025competition,gupta2023centralizing} show that today’s builder market is highly centralized. In Ethereum, two builders have been producing 85--90\% of blocks. The concentration has led to significant proposer losses. Many block proposers do not receive the full MEV value because dominant builders only bid slightly above the next-best bids. Moreover, private order flow~\cite{wang2025private,lyu2025demystifying,zhang2024breaking} makes things worse. Top builders gain access to private transaction order flow (from searchers or order-flow providers) that others cannot, creating a feedback loop where market share begets more exclusive flow. This centralization is even more pronounced in Binance’s PoSA setting. The small validator set and shorter block intervals amplify builders’ structural advantages.

\vspace{3pt}
\noindent\textbf{Binance's PoSA.} BSC adopts PoSA \cite{bep126,li2025finality}, which organizes time into epochs and slots (3s per slot). In each epoch, a small committee of validators, elected via stake delegation, takes turns proposing and sealing blocks. Because validator turnover is infrequent due to high staking costs, the set remains relatively static.

The resulting short block interval limits the time window available for searchers/builders to capture MEV. This has two implications: it may constrain complex cross-pool or cross-chain arbitrage, while simultaneously amplifying the advantages of builders with low-latency infrastructure or privileged order flow. How PoSA’s fast-paced consensus interacts with a builder market remains an open question that deserves fuller causal analysis in future work.

%=================================================
\section{Conclusion}
\label{sec-conclu}
%=================================================

Binance’s PBS is more centralized than Ethereum in both participation scale and block production dominance with only two builders producing 87\%+ of blocks. The centralization enables builders to capture nearly all MEV revenues while \bd{48Club} alone captures about three quarters of observed net profit.  Binance also shows weaker
censorship resistance, reduced structural fairness, and diminished
MEV competition. We empirically demonstrate these claims.

%================================================
%================================================
\bibliographystyle{unsrt}
\bibliography{bib.bib}
%================================================
%================================================

\appendix

%=================================
\section{Preparing Data}
\label{apd-data}
%=================================

This appendix expands the data collection and processing steps. We cover raw trace acquisition, builder-contract labeling, arbitrage-cycle reconstruction, and profit attribution.

%-----------------
\subsection{Extended Data Collection (\S\ref{subsec:data-collection}, \S\ref{subsec:arbitrage})}
\label{apd-method}
%-----------------

This appendix expands the technical details underlying our data collection and processing.

\vspace{3pt}
\noindent\textbf{Raw data acquisition.}
We collect raw blockchain data directly from BSC full nodes over the block
ranges associated with \bd{48Club} and \bd{Blockrazor}.  
Our crawler retrieves:
\begin{packeditemize}
    \item block headers, full block bodies, and transaction receipts;
    \item per-transaction execution traces via the client tracing API;
    \item contract bytecode and emitted events for all addresses interacting with builder-linked contracts.
\end{packeditemize}
These artifacts form the basis for reconstructing swap paths, liquidity-pool
interactions, and profit flows.

\vspace{3pt}
\noindent\textbf{Path reconstruction and arbitrage identification.}
For each transaction initiated by a builder-owned contract, we rebuild its full
execution path by parsing \texttt{Swap}, \texttt{Sync}, \texttt{Transfer}, and
router-level events.  
A transaction is labeled as an arbitrage cycle if:
\[
    \mathsf{asset}_{\mathrm{in}} = \mathsf{asset}_{\mathrm{out}},
\]
with multi-hop sequences treated as a single atomic opportunity.  
We further annotate path length, intermediate pools, and routing category
(stablecoins, wrapped assets, or mixed).

\vspace{3pt}
\noindent\textbf{Profit attribution rules.}
For each arbitrage we compute:
\begin{packeditemize}
    \item \emph{Gross profit}: raw output minus input (in token units),
    \item \emph{Share profit}: any portion redirected to validator-revenue endpoints (e.g., \texttt{0xffff...fffe} observed in our traces; see §4.2),
    \item \emph{Net profit}: builder-retained profit after subtracting share-profit and gas cost.
\end{packeditemize}

And the special handling is required for:
\begin{packeditemize}
    \item \emph{Pool-routed profits}: surplus tokens routed directly into pools;
    \item \emph{Intermediary-token dust}: negligible residuals left inside intermediary contracts.
\end{packeditemize}

\vspace{3pt}
\noindent\textbf{Gas accounting.}
Gas consumption was extracted from receipts. We observed near-zero gas prices (0 Gwei) in most arbitrage transactions, consistent with the design of BSC PBS where builders compete via bids rather than gas fees. This feature shows the role of PBS in decoupling transaction execution from direct fee incentives.

\vspace{3pt}
\noindent\textbf{Processing pipeline.}
Given dataset scale, we:
\begin{packeditemize}
    \item deduplicate multi-route opportunities,
    \item normalize profit values into BNB-equivalent amounts using a fixed reference \texttt{WBNB} price, and
    \item aggregate per-builder/per-token/per-path statistics,
    \item remove malformed traces.
\end{packeditemize}

Each arbitrage entry is annotated with the base token, path length, pools used, gross/share/net profits, and redistribution endpoints. Random samples are manually checked by BscScan.

\begin{algorithm}[H]
\footnotesize
\caption{ExtractArbitrageCycle$(tx)$}
\label{algo-extractArbi}
\begin{algorithmic}[1]
\Require trace events $T$ of transaction $tx$
\Ensure arbitrage cycle $C$ or $\bot$

\State $P \gets [\,]$

\For{each event $e$ in $T$}
    \If{$e.\mathrm{type} = \textsc{Swap}$} 
        \State append $(e.\mathrm{token\_in}, e.\mathrm{token\_out}, e.\mathrm{pool})$ to $P$
    \EndIf
\EndFor

\If{$P = \emptyset$}        
    \State \Return $\bot$
\EndIf

\State $(a_{\mathrm{in}},\,\_) \gets P[0]$
\State $(\_,\,a_{\mathrm{out}}) \gets P[\mathrm{last}]$

\If{$a_{\mathrm{in}} \ne a_{\mathrm{out}}$}
    \State \Return $\bot$
\EndIf

\State \Return $P$
\end{algorithmic}
\end{algorithm}

\vspace{3pt}
\noindent\textbf{Validator’s income via \texttt{0xffff...fffe}.}
A recurring artifact is transfers to \texttt{0xffff...}, which builders use in practice to forward validator-side revenue. 

We attribute these flows as follows:
\begin{packeditemize}
    \item extracting block proposer from consensus metadata,
    \item linking transfers to \texttt{0xffff...} in that block to validator,
    \item grouping ambiguous cases under ``Unknown/Other''.
\end{packeditemize}

Our empirical study attributes over 95\% of flows; the remainder is negligible.

%-----------------
\subsection{Full Example and Algorithm}
\label{apd-example}
%-----------------

We show a simple trace excerpt from a \bd{48Club} arbitrage:

\begin{lstlisting}[style=trace]
Trace[0]  Swap(USDT -> WBNB, pool=0xA1..., amt_in=1,000,000, amt_out=2.98)
Trace[1]  Swap(WBNB -> USD1, pool=0xB4..., amt_in=2.98,      amt_out=1,001,120)
Trace[2]  Swap(USD1 -> USDT, pool=0xC7..., amt_in=1,001,120, amt_out=1,003,040)
Trace[3]  Transfer(to=0xffff...fffe, amt=820)      // validator share-profit
Trace[4]  InternalTxn(to=builder_wallet, amt=1,920) // retained net profit
\end{lstlisting}

Reconstruction yields:
\begin{compactitem}
    \item Base token: USDT
    \item Path: USDT $\rightarrow$ WBNB $\rightarrow$ USD1 $\rightarrow$ USDT
    \item Gross profit: 3{,}040 units
    \item Share-profit: 820 units
    \item Net profit: 2{,}220 units (BNB-equivalent in analysis)
\end{compactitem}

\medskip
\noindent\textbf{Arbitrage-cycle extraction} (Algorithm~\ref{algo-extractArbi}).
This procedure recovers the arbitrage cycle executed by a builder contract by
scanning all swap-related events in a transaction trace. It outputs an ordered
swap sequence when the entry and exit assets match.

\medskip
\noindent\textbf{Profit attribution} (Algorithm~\ref{algo-attributeprofit}).
This procedure computes the three components of builder revenue for an
arbitrage cycle: gross profit, redistributed share-profit (e.g., validator
income), and final net profit. It aggregates explicit transfers as well as
implicit pool-routed profits.

\begin{algorithm}[H]
\footnotesize
\caption{AttributeProfit$(tx, P)$}
\label{algo-attributeprofit}
\begin{algorithmic}[1]
\Require transaction $tx$, arbitrage path $P$
\Ensure $(\mathrm{gross},\,\mathrm{share},\,\mathrm{net})$

\State $v_{\mathrm{in}}  \gets$ input amount of base token
\State $v_{\mathrm{out}} \gets$ output amount of base token
\State $\mathrm{gross} \gets v_{\mathrm{out}} - v_{\mathrm{in}}$

\State $\mathrm{share} \gets 0$

\For{each event $e$ in $tx$}
    \If{$e.\mathrm{type}=\textsc{Transfer}$ and $e.\mathrm{to}\in\{\texttt{0xffff...},~\text{sysAddrs}\}$}
        \State $\mathrm{share} \gets \mathrm{share} + e.\mathrm{amt}$
    \ElsIf{$e.\mathrm{type}=\textsc{Swap}$ and $e.\mathrm{pool\_sink}=\mathrm{true}$}
        \State $\mathrm{share} \gets \mathrm{share} + e.\mathrm{amt\_routed}$
    \EndIf
\EndFor

\State $\mathrm{gas} \gets tx.\mathrm{gas\_used} \times tx.\mathrm{gas\_price}$

\State $\mathrm{net} \gets \mathrm{gross} - \mathrm{share} - \mathrm{gas}$

\State \Return $(\mathrm{gross},\mathrm{share},\mathrm{net})$
\end{algorithmic}
\end{algorithm}

%-----------------
\subsection{Builder List (cf. \S\ref{subsec-builder})}
\label{apd-builderlist}
%-----------------

\begin{table*}[ht]
\centering
\footnotesize
\renewcommand{\arraystretch}{1.2}
\caption{Binance builder list over the April~1,~2025 to February~28,~2026 observation window}
\label{tab:builder}
%\resizebox{\textwidth}{!}{
\begin{tabular}{crccc}

\toprule
\makecell{\textbf{Builder Brand }} & \multicolumn{1}{c}{\textbf{Builder Name \& Address} } & \quad \textbf{Blocks} \quad  & \quad \makecell{\textbf{Integrated} \textbf{Validators}} \quad  & \quad \makecell{\textbf{Market} \textbf{Share}} \quad  \\

\midrule

\hlhref{https://www.48.club}{\bd{48Club-puissant}} & 
\quad\quad\quad \makecell{
(48Club-puissant-1) \addr{0x487e5d…bb48}{https://bscscan.com/address/0x487e5dfe70119c1b320b8219b190a6fa95a5bb48} \\ 
(48Club-puissant-2) \addr{0x48a5ed…bb48}{https://bscscan.com/address/0x48a5ed9abc1a8fbe86cec4900483f43a7f2dbb48}\\
(48Club-puissant-3) \addr{0x48fee1…bb48}{https://bscscan.com/address/0x48fee1bb3823d72fdf80671ebad5646ae397bb48}\\
(48Club-puissant-4)  \addr{0x48b4bb…bb48}{https://bscscan.com/address/0x48b4bbebf0655557a461e91b8905b85864b8bb48}\\
(48Club-puissant-5)  \addr{0x48b266…bb48}{https://bscscan.com/address/0x48b2665e5e9a343409199d70f7495c8ab660bb48}\\
(48Club-puissant-6)  \addr{0x4827b4…bb48}{https://bscscan.com/address/0x4827b423d03a349b7519dda537e9a28d31ecbb48}
}\quad\quad\quad & \quad\quad 17,965,461 \quad\quad & 45 & 46.57\% \\

\midrule
\hlhref{https://www.blockrazor.io/}{\bd{Blockrazor}} & \makecell{
(Blockrazor: Builder1) \addr{0x5532cd…7455}{https://bscscan.com/address/0x5532cdb3c0c4278f9848fc4560b495b70ba67455}\\
(Blockrazor: Builder2)  \addr{0x49d91b…6149}{https://bscscan.com/address/0x49d91b1ab0cc6a1591c2e5863e602d7159d36149}\\
(Blockrazor: Builder3)  \addr{0xba4233…7be1}{https://bscscan.com/address/0xba4233f6e478db76698b0a5000972af0196b7be1}\\
(Blockrazor: Builder4) \addr{0x539e24…5c53}{https://bscscan.com/address/0x539e24781f616f0d912b60813ab75b7b80b75c53}\\
(Blockrazor: Builder5)  \addr{0x500610…e250}{https://bscscan.com/address/0x50061047b9c7150f0dc105f79588d1b07d2be250}\\
(Blockrazor: Builder6)  \addr{0x488e37…6448}{https://bscscan.com/address/0x488e37fcb2024a5b2f4342c7de636f0825de6448}\\
(Blockrazor: Builder7)  \addr{0x0557e8…a597}{https://bscscan.com/address/0x0557e8cb169f90f6ef421a54e29d7dd0629ca597}
} & 15,916,288 & 45 & 41.26\% \\

\midrule
\hlhref{https://jetbldr.xyz/}{\bd{Jetbldr}} & \makecell{
(Builder: jetbldr-1) \addr{0x345324…3b4c}{https://bscscan.com/address/0x345324dc15f1cdcf9022e3b7f349e911fb823b4c}\\
(Builder: jetbldr-2) \addr{0x36cb52…ebb5}{https://bscscan.com/address/0x36cb523286d57680efbbfb417c63653115bcebb5}\\
(Builder: jetbldr-3) \addr{0x3ad612…06a8}{https://bscscan.com/address/0x3ad6121407f6edb65c8b2a518515d45863c206a8}\\
(Builder: jetbldr-4) \addr{0xfd3835…dca1}{https://bscscan.com/address/0xfd38358475078f81a45077f6e59dff8286e0dca1}\\
(Builder: jetbldr-5) \addr{0x7f5fbf…69a3}{https://bscscan.com/address/0x7f5fbfd8e2eb3160df4c96757deef29e26f969a3}\\
(Builder: jetbldr-6) \addr{0xa0cde9…9ab43}{https://bscscan.com/address/0xa0cde9891c6966fce740817cc5576de2c669ab43}
} & 991,850 & 40 & 2.57\% \\

\midrule

\hlhref{https://bloxroute.com/}{\bd{Bloxroute}} & \makecell{
(Bloxroute: Builder1) \addr{0xd4376f…fd52}{https://bscscan.com/address/0xd4376fdc9b49d90e6526daa929f2766a33bffd52}\\
(Bloxroute: Builder2) \addr{0x2873fc…388d}{https://bscscan.com/address/0x2873fc7ad9122933becb384f5856f0e87918388d}\\
(Bloxroute: Builder3) \addr{0x432101…1c66}{https://bscscan.com/address/0x432101856a330aafdeb049dd5fa03a756b3f1c66}\\
(Bloxroute: Builder4) \addr{0x2b217a…3ae7}{https://bscscan.com/address/0x2b217a4158933aade6d6494e3791d454b4d13ae7}\\
(Bloxroute: Builder5) \addr{0x0da52e…6415}{https://bscscan.com/address/0x0da52e9673529b6e06f444fbbed2904a37f66415}\\
(Bloxroute: Builder6) \addr{0xe1ec1a…860a}{https://bscscan.com/address/0xe1ec1aece7953ecb4539749b9aa2eef63354860a}\\
(Bloxroute: Builder7) \addr{0x89434f…6a15}{https://bscscan.com/address/0x89434fc3a09e583f2cb4e47a8b8fe58de8be6a15}\\
(Bloxroute: Builder8) \addr{0x103535…74ff}{https://bscscan.com/address/0x10353562e662e333c0c2007400284e0e21cf74ff}
} & 719,508 & 44 & 1.86\% \\

\midrule
%\hdashline[1pt/1pt]
\hlhref{https://nodereal.io/}{\bd{Nodereal}} & \makecell{
(Nodereal: Builder1) \addr{0x79102d…8fa0}{https://bscscan.com/address/0x79102db16781dddff63f301c9be557fd1dd48fa0}\\
(Nodereal: Builder2) \addr{0xd0d56b…1b6f}{https://bscscan.com/address/0xd0d56b330a0dea077208b96910ce452fd77e1b6f}\\
(Nodereal: Builder3) \addr{0x4f24ce…0b99}{https://bscscan.com/address/0x4f24ce4cd03a6503de97cf139af2c26347930b99}\\
(Nodereal: Builder4) \addr{0x5b526b…6c48}{https://bscscan.com/address/0x5b526b45e833704d84b5c2eb0f41323da9466c48}\\
(Nodereal: Builder5) \addr{0xa547f8…605e}{https://bscscan.com/address/0xa547f87b2bade689a404544859314cbc01f2605e}\\
(Nodereal: Builder6) \addr{0xfd3f1a…7335}{https://bscscan.com/address/0xfd3f1ad459d585c50cf4630649817c6e0cec7335}
} & 383,758 & 36 & 0.99\% \\

\midrule

\hlhref{https://www.blocksmithlabs.io/}{\bd{Blocksmith}} & \makecell{
(Blocksmith: Builder1) \addr{0x6dddf6…e5c7}{https://bscscan.com/address/0x6dddf681c908705472d09b1d7036b2241b50e5c7}\\
(Blocksmith: Builder2) \addr{0x767361…8191}{https://bscscan.com/address/0x76736159984ae865a9b9cc0df61484a49da68191}\\
(Blocksmith: Builder3) \addr{0x5054b2…23e}{https://bscscan.com/address/0x5054b21d8baea3d602dca8761b235ee10bc0231e}
} & 147,228 & 35 & 0.38\% \\

\bottomrule
\end{tabular}
%}
\end{table*}

We obtained the list of active builders (Table~\ref{tab:builder}) from BSC’s
official GitHub repository (\url{https://github.com/bnb-chain/bsc-mev-info/blob/main/mainnet/builder-list.toml})
and cross-checked it with Dune (\url{https://dune.com/bnbchain/bnb-smart-chain-mev-stats}).

\bd{48Club} is one of the earliest whitelisted builders and the most dominant,
operating multiple endpoints across validators and capturing the majority of
private order flow.

\bd{Blockrazor} is a secondary but consistently active builder operating fewer
endpoints with narrower arbitrage coverage, frequently appearing in validator
logs and MEV dashboards.

%=================================================
%\section{Toy Extension: Arbitrage Abstraction and Instance}
%\label{sec-extenstion}
%=================================================

%As an independent line of interest, we further explore how arbitrageurs can be abstracted into a generalized framework. 
%Rather than committing to a single implementation, we aim to capture the conceptual structure: 
%arbitrageurs maintain a set of personalized strategies, apply them to transaction streams to identify profitable opportunities, and may adopt additional \textit{evasion steps} (forbidden in \cite{bartoletti2023theoretical}) such as routing through private relays or obscuring transaction patterns to protect their methods from replication or frontrunning. 

%We provide a concrete algorithmic implementation of this abstraction, which is deferred to Appendix~\ref{apd-imple}.  We hope to raise community awareness and encourage further discussion of MEV arbitrage mechanisms.

\begin{algorithm}[t]
\footnotesize
\caption{ArbitrageRun — Instantiated Mixed-Path Execution}
\label{alg:arb-run}
\begin{algorithmic}[1]
\Require Strategy $s \in \mathcal{S}$; transaction stream $\mathcal{T}$; opportunity $O=f_s(\mathcal{T})$; optional evasion operator $\mathcal{E}$
\Ensure Profit returned in \texttt{WBNB} at contract; share payout; retained remainder
\State $(\langle T_0,\ldots,T_n\rangle, \langle P_1,\ldots,P_n\rangle, \langle F_1,\ldots,F_n\rangle, \langle D_1,\ldots,D_n\rangle) \gets \mathcal{E}(O)$
\State Assert $\#tokens=n{+}1$, $\#pools=n$, $\#flags=n$, $\#dirs=n$ \label{line:assert}
\State Transfer $A_0$ of $T_0$ from user to contract \label{line:transfer-init}
\State $B_{\text{pre}}\gets \mathrm{bal}(\texttt{WBNB},\text{contract})$;\quad $A\gets A_0$ \label{line:init-balance}
\For{$i=1$ to $n$} \label{line:forloop}
  \State $(T_{\text{in}},T_{\text{out}},P)\gets (T_{i-1},T_i,P_i)$
  \State $\text{rcv}\gets \text{contract if }i{=}n\text{ else }P_{i+1}$ \label{line:recipient}
  \State Transfer $A$ of $T_{\text{in}}$ to $P$ \label{line:transfer-hop}
  \If{$F_i=0$} \Comment{V3 exact-input} \label{line:v3-if}
    \State $P.\texttt{swap}(\text{rcv}, D_i, +A, \mathrm{PriceLimit}(D_i), \emptyset)$ \label{line:v3-swap}
  \Else \Comment{V2} \label{line:v2-else}
    \State Compute $(a_0,a_1)$ from $D_i$; call $P.\texttt{swap}(a_0,a_1,\text{rcv},\emptyset)$ \label{line:v2-swap}
  \EndIf
  \State $A\gets \mathrm{bal}(T_{\text{out}},\text{rcv})$ \label{line:update-amount}
\EndFor
\If{$P^\star\neq\varnothing$} \Comment{optional final hop to WBNB} \label{line:final-if}
  \State Transfer $A$ of $T_n$ to $P^\star$
  \State $P^\star.\texttt{swap}(\text{contract}, D^\star, +A, \mathrm{PriceLimit}(D^\star), \emptyset)$ \label{line:final-swap}
\EndIf
\State $B_{\text{post}}\gets \mathrm{bal}(\texttt{WBNB},\text{contract})$;\quad \textbf{require }$B_{\text{post}}>B_{\text{pre}}$ \label{line:profit-check}
\State $\Delta\gets B_{\text{post}}-B_{\text{pre}}$;\quad $\text{payout}\gets \lfloor \Delta\cdot r/10{,}000\rfloor$ \label{line:calc-profit}
\State Unwrap \text{payout} \texttt{WBNB} $\to$ BNB;\; transfer to \texttt{0xffff...fffe} \label{line:unwrap}
\State \Return kept \texttt{WBNB} profit $= \Delta - \text{payout}$ \label{line:return}
\end{algorithmic}
\end{algorithm}

%====================================================
\section{Implement Instance (recall \S\ref{sec-bck})}
\label{apd-imple}
%====================================================

We explore how arbitrageurs can be abstracted into a generalized framework.  Aligned with \S\ref{sec-bck}, here is a conceptual structure: arbitrageurs maintain a set of personalized strategies, apply them to transaction streams to identify profitable opportunities, and may adopt additional \textit{evasion steps} (forbidden in \cite{bartoletti2023theoretical}) such as routing through private relays or obscuring transaction patterns to protect their methods from replication or frontrunning.

\vspace{3pt}
\noindent\textbf{Execution logic} (Algorithm~\ref{alg:arb-run}). This template shows how hidden strategies $\mathcal{S}$ and optional evasion operators $\mathcal{E}$ are integrated into concrete arbitrage executions on BSC.

The contract begins by parsing the descriptor $(T,P,F,D)$ after applying any evasion operator $\mathcal{E}$ (Alg.~\ref{alg:arb-run}, line~1) and checks consistency (line~\ref{line:assert}).  
It then transfers the initial token amount $A_0$ into the contract (line~\ref{line:transfer-init}) and records the baseline \texttt{WBNB} balance (line~\ref{line:init-balance}).  

For each hop (line~\ref{line:forloop}), the current token amount is forwarded to the corresponding pool $P_i$ (line~\ref{line:transfer-hop}), where either a V3 exact-input swap (lines~\ref{line:v3-if}--\ref{line:v3-swap}) or a V2 swap (lines~\ref{line:v2-else}--\ref{line:v2-swap}) is executed, and the output becomes the input for the next hop (line~\ref{line:update-amount}).  
If a final normalization hop $P^\star$ exists, the proceeds are swapped back to \texttt{WBNB} (lines~\ref{line:final-if}--\ref{line:final-swap}).  
Finally, the contract checks profitability (line~\ref{line:profit-check}), computes the net surplus $\Delta$ and splits it according to share ratio $r$ (line~\ref{line:calc-profit}), unwrapping the payout to BNB for validator distribution (line~\ref{line:unwrap}) while retaining the remainder as realized profit (line~\ref{line:return}).  

\vspace{3pt}
\noindent\textbf{A brief comparison.} We note that prior formal models of MEV~\cite{bartoletti2023theoretical} rule out private order flow by assuming a “no shared secrets” setting. This assumption fits Ethereum’s PBS, where builders compete in an open relay market and all opportunities are assumed visible in the public mempool. 

Binance’s PBS, however, works differently. Private channels are the norm: builders and validators connect through whitelisted APIs, users and searchers send bundles via private RPCs, and 0~Gwei transactions largely bypass the public mempool. To reflect this reality, our abstraction includes an optional evasion operator $\mathcal{E}$, which captures how opportunities can be reshaped or concealed before execution. 
We conclude that Ethereum-based models see private flow as an exception, but on Binance, it should be placed as default and treated as part of the standard arbitrage workflow.

\end{document}